\documentclass[12pt]{article}
\RequirePackage[OT1]{fontenc}
\usepackage{amsmath,amssymb,amsthm}
\usepackage{natbib}
\usepackage{graphicx}
\usepackage{bm}

\usepackage[pdftex,colorlinks=true,linkcolor=blue,citecolor=blue,urlcolor=blue,bookmarks=false,pdfpagemode=None]{hyperref}
\usepackage{verbatim}
\usepackage[top=1in, bottom=1in, left=1in, right=1in]{geometry}
\usepackage{setspace}
\usepackage{array}
\usepackage{rotating}
\usepackage{float}
\usepackage{enumitem}
\usepackage{xcolor}
\usepackage{algorithm2e}
\usepackage{framed}
\usepackage{threeparttable}
\usepackage{appendix}
\usepackage{multirow}
\usepackage{geometry}
\usepackage{url}
\usepackage{authblk}

\newcommand\independent{\protect\mathpalette{\protect\independenT}{\perp}}
\def\independenT#1#2{\mathrel{\rlap{$#1#2$}\mkern2mu{#1#2}}}

\newcolumntype{L}[1]{>{\raggedright\arraybackslash}p{#1}}
\RequirePackage{subfigure}
\RequirePackage{graphicx}
\RequirePackage{xspace}
\RequirePackage{tabularx}
\RequirePackage{comment}
\RequirePackage{subdepth}

\usepackage{fancyhdr}
\usepackage{tikz}
\usetikzlibrary{matrix,fit,backgrounds}
\usetikzlibrary{decorations.pathreplacing}

\usepackage{color}

\DeclareSymbolFont{AMSb}{U}{msb}{m}{n}
\DeclareMathSymbol{\N}{\mathbin}{AMSb}{"4E}
\DeclareMathSymbol{\Z}{\mathbin}{AMSb}{"5A}
\DeclareMathSymbol{\R}{\mathbin}{AMSb}{"52}
\DeclareMathSymbol{\Q}{\mathbin}{AMSb}{"51}
\DeclareMathSymbol{\I}{\mathbin}{AMSb}{"49}
\DeclareMathSymbol{\C}{\mathbin}{AMSb}{"43}
\DeclareMathSymbol{\D}{\mathbin}{AMSb}{"44}
\DeclareMathSymbol{\E}{\mathbin}{AMSb}{"45}

\def\b1{\boldsymbol{1}}

\def\RR{\hbox{I\kern-.1667em\hbox{R}}}

\newcommand{\EE}[1]{\E \left [ #1 \right ]}

\begin{document}

\def\spacingset#1{\renewcommand{\baselinestretch}%
{#1}\small\normalsize} \spacingset{1}
 
  \title{\bf Matching with Text Data: An Experimental Evaluation of Methods for Matching Documents and of Measuring Match Quality}
  \author[1]{Reagan Mozer\thanks{Corresponding Author: Reagan Mozer (email: rmozer@g.harvard.edu)}}
\author[1]{Luke Miratrix}
\author[1]{Aaron Russell Kaufman}
\author[2]{L. Jason Anastasopoulos}
\affil[1]{Harvard University}
\affil[2]{University of Georgia}

\maketitle

\bigskip
\begin{abstract}
Matching for causal inference is a well-studied problem, but standard methods fail when the units to match are text documents: the high-dimensional and rich nature of the data renders exact matching infeasible, causes propensity scores to produce incomparable matches, and makes assessing match quality difficult. In this paper, we characterize a framework for matching text documents that decomposes existing methods into: (1) the choice of text representation, and (2) the choice of distance metric. We investigate how different choices within this framework affect both the quantity and quality of matches identified through a systematic multifactor evaluation experiment using human subjects. Altogether we evaluate over 100 unique text matching methods along with 5 comparison methods taken from the literature. Our experimental results identify methods that generate matches with higher subjective match quality than current state-of-the-art techniques. We enhance the precision of these results by developing a predictive model to estimate the match quality of pairs of text documents as a function of our various distance scores. This model, which we find successfully mimics human judgment, also allows for approximate and unsupervised evaluation of new procedures in our context. We then employ the identified best method to illustrate the utility of text matching in two applications. First, we engage with a substantive debate in the study of media bias by using text matching to control for topic selection when comparing news articles from thirteen news sources. We then show how conditioning on text data leads to more precise causal inferences in an observational study examining the effects of a medical intervention.
\end{abstract}

\vfill


\newpage
\spacingset{1.5}

\section{Introduction}
\label{sec:intro}
Recently, \cite{roberts2018adjusting} introduced an approach for matching text documents in order to address confounding in observational studies of substantive and policy-relevant quantities of interest. Matching is a statistical tool primarily used to facilitate causal inferences about the effects of a particular treatment, action, or intervention from non-randomized data in the presence of confounding covariates \citep{rubin1973use, rosenbaum2002observational, rubin2006matched, stuart2010matching}. The principles behind matching can also be used to create sharp, targeted comparisons of units in order to, for example, create more principled rankings of hospitals \citep{silber2014template}. The core idea of matching is to find sets of units from distinct populations that are in all ways similar, other than some specific aspects of interest; one can then compare these remaining aspects across the populations of interest to ascertain differences foundational to these populations. In short, matching provides a strategy for making precise comparisons and performing principled investigations in observational studies. 

Though widely used in practice, matching is typically used in settings where both the covariates and outcomes are well-defined, low-dimensional quantities. Text is not such a setting. With text, standard contrasts of outcomes between groups may be distorted estimates of the contrasts of interest due to confounding by high-dimensional and possibly latent features of the text such as topical content or overall sentiment. How to best capture and adjust for these features is the core concern of this work. In particular, we consider the problem of matching documents within a corpus made up of distinct groups (e.g., a treatment and control group), where interest is in finding a collection of matched documents that are fundamentally ``the same'' along key dimensions of interest (in our first application, for example, we find newspaper articles that are about the same events and stories). These matched documents can then be used to make unbiased comparisons between groups on external features such as rates of citation or online views, or on features of the text itself, such as sentiment. In the case where group membership can be thought of as the receipt of a particular intervention (e.g., documents that were censored vs. not, such as in \citealp{roberts2018adjusting}), this allows us draw causal inferences about effects of interest.

This paper makes three contributions to guide researchers interested in this domain. Our first contribution is a deconstruction and discussion of the elements that constitute text matching. This formulation identifies a series of choices a researcher can make when performing text matching and presents an approach for conceptualizing how matching can be used in studies where the covariates, the outcome of interest, or both are defined by summary measures of text. Our second contribution is to investigate these choices using a systematic multifactor human evaluation experiment to examine how different representations and distance metrics correspond to human judgment about document similarity. Our experiment explores the efficiency of each combination of choices for matching documents in order to identify the representations and distance metrics that dominate in our context in terms of producing the largest number of matches for a given dataset without sacrificing match quality. We also present a general framework for designing and conducting systematic evaluations of text-matching methods that can be used to perform similar investigations in different contexts. Our third contribution is twofold. 

First, we present a novel application of template matching \citep{silber2014template} to compare news media organizations' biases, beyond choices of which stories to cover, in order to engage with a running debate on partisan bias in the news media. Through template matching on text, we identify similar samples of news articles from each news source that, taken together, allow for a more principled (though not necessarily causal) investigation of how different news sources may differ systematically in terms of partisan favorability. In our second application, we illustrate the utility of text matching in a more traditional causal inference setting, namely, in an observational study evaluating the causal effects of a binary treatment. Here we demonstrate how matching on text obtained from doctors' notes can be used to improve covariate balance between treatment and control groups in an observational study examining the effects of a medical intervention. We further discuss how researchers might leverage text data to strengthen the key assumptions required to make valid causal inferences in this non-randomized context.

Our work builds on \cite{roberts2018adjusting}, the seminal paper in this literature, which introduces text matching, operationalizing the text data by using topic modeling coupled with propensity scores to generate a lower-dimensional representation of text to match on. They also present several applications that motivate the use of text matching to address confounding and describe several of the methodological challenges for matching that arise in these settings. Specifically, \cite{roberts2018adjusting} discuss the limitations of direct propensity score matching and coarsened exact matching (CEM) on the raw text for matching with high dimensional data and introduce Topical Inverse Regression Matching (TIRM), which uses structural topic modeling (STM) \citep{roberts2016model} to generate a low-dimensional representation of a corpus and then applies CEM to generate matched samples of documents from distinct groups within the corpus. Building upon this work, we develop a general framework for constructing and evaluating text matching methods. This allows us to consider a number of alternative matching methods not considered in \cite{roberts2018adjusting}, each characterized by one representation of the corpus and one distance metric. Within this framework, we also present a systematic approach for comparing different matching methods through our evaluation experiment, which identifies methods that can produce more matches and/or matches of higher quality than those produced by TIRM. Overall, we clarify that there is a tradeoff between match quality and the number of matches, although many methods do not optimize either choice.

\section{Background}
\label{sec:background}

\subsection{Notation and problem setup}
Consider a collection of $N$ text documents, indexed by $i=1,\ldots,N$, where each document contains a sequence of terms. These documents could be any of a number of forms such as news articles posted online, blog posts, or entire books, and each document in the dataset need not be of the same form. Together, these $N$ documents comprise a corpus, and the set of $V$ unique terms used across the corpus define the vocabulary. Each term in the vocabulary is typically a unique, lowercase, alphanumeric token (i.e., a word, number, or punctuation mark), though the exact specification of terms may depend on design decisions by the analyst (e.g., one may choose to include as terms in the vocabulary all bigrams observed in the corpus in addition to all observed unigrams). Because the number and composition of features which may be extracted from text is not well defined, documents are generally regarded as ``unstructured'' data in the sense that their dimension is \textit{ex ante} unknown.\footnote{In particular, the number and composition of features which may be extracted from a given corpus is not well-defined and may vary depending on researcher focus.} 
To address this issue, we impose structure on the text through a representation, $X$, which maps each document to a finite, usually high-dimensional, quantitative space. 

To make principled comparisons between groups of documents within the corpus, we borrow from the notation and principles of the Rubin Causal Model (RCM) \citep{holland1986statistics}. 
Under the RCM, each document has an indicator for treatment assignment (i.e., group membership), $Z_i$, which equals 1 for documents in the treatment group and 0 for documents in the control group. Interest focuses on estimating differences between these groups on an outcome variable, which, under a causal view, would take the value $Y_i(1)$ if document $i$ is in the treatment group and $Y_i(0)$ if document $i$ is in the control group. 
These outcomes may be separate from the text of the document (e.g., the number of times a document has been viewed online) or may be a feature of the text (e.g., the length of the document or level of positive sentiment within the document).\footnote{In the latter case, care must be taken to ensure the features of the representation $X$ used to define the covariates are suitably separated from features that define the potential outcomes. This issue is discussed further in Section~\ref{sec:methods} and in Appendix~\ref{app:covar}.} 
Credible and precise causal inference revolves around comparing treated and control documents that are as similar as possible. 
However, in observational studies, $Z_i$ is typically not be randomly assigned, leading to systematic differences between treatment and control groups. 
Matching is a strategy that attempts to address this issue by identifying samples of treated and control documents that are comparable on covariates in order to approximate random assignment of $Z_i$ (i.e., to satisfy $Z_i \independent (Y_i(0),Y_i(1)) | X_i$) \citep{rosenbaum2002observational, rubin2006matched}. 
Under this key assumption of ``selection on observables,'' which states that all covariates that affect both treatment assignment and potential outcomes are observed and captured within $X$, comparisons of outcomes between matched samples can be used to obtain unbiased estimates of the quantities of interest \citep{rosenbaum2002observational}.
For example, in our second application examining the effects of a medical intervention, we argue that matching on both a set of numerical covariates and the text content of the patients chart allows us to identify two groups of patients, one treated and one not, that are similar enough on pre-treatment variables such that any systematic differences in their outcomes can be plausibly attributed to the impact of the intervention.

These causal inference tools can be used more broadly, however, to produce clearly defined comparisons of groups of units even when a particular intervention is not well-defined. For example, \cite{silber2014template} introduces \emph{template matching} as a tool for comparing multiple hospitals that potentially serve different mixes of patients (e.g., some hospitals have a higher share of high-risk patients). The core idea is to compare like with like: by comparing hospitals along an effective ``score card'' of patients, we can see which hospitals are more effective, on average, given a canonical population. In general, we focus on this general conception of matching, recognizing that often in text there is no treatment that could, even in concept, be randomized. For example, a comparison of style between men and women could not easily be construed as a causal impact. Nevertheless, the framing and targeting of a controlled comparison, a framing inherent in a causal inference approach, can still be useful in these contexts. This broader formulation of matching is used in our first application in Section~\ref{sec:applied} investigating different aspects of bias in newspaper media.

\subsection{Promises and pitfalls of text matching}
Matching methods generally consist of five steps: 1) identify a collection of potential confounders (covariates) that would compromise any causal claims if they were systematically different across the treatment groups, 2) define a measure of distance (or similarity) to determine whether one unit is a good match for another, 3) match units across groups according to the chosen distance metric, 4) evaluate the quality of the resulting matched samples in terms of their balance on observed covariates, possibly repeating the matching procedure until suitable balance is achieved, and 5) estimate treatment effects from these matched data \citep{stuart2010matching}. 
Different choices at each step of this process produce an expansive range of possible configurations. For instance, there are distance metrics for scalar covariates \citep{rubin1973use}, for multivariate covariates summarized through a univariate propensity score \citep{rosenbaum1983central, rosenbaum1985constructing}, and multivariate metrics such as the Mahalanobis distance metric \citep{rubin1978bias, gu1993comparison}. 

Similarly, there is a large and diverse literature on matching procedures \citep{rosenbaum2002observational, rubin2006matched}, and the choice of procedure depends on both substantive and methodological concerns. 
Some procedures match each unit in the treatment group to its one ``closest'' control unit and discard all unused controls (e.g., one-to-one matching with replacement), while other procedures allow treated units to be matched to multiple controls (e.g., ratio matching; \citealp{smith1997matching}) and/or matching without replacement (e.g., optimal matching; \citealp{rosenbaum1989optimal}). 
Match quality is often evaluated with a number of diagnostics that formalize the notion of covariate balance such as the standardized differences in means of each covariate \citep{rosenbaum1985constructing}. 
Unfortunately, determinations of what constitutes ``suitable'' balance or match quality are often based on arbitrary criteria \citep{imai2008misunderstandings, austin2009balance}, and assessing whether a matching procedure has been successful can be quite difficult. 
That being said, once a suitable set of matches is obtained, one can then typically analyze the resulting matched data using classic methods appropriate for the type of data in hand.  
\cite{stuart2010matching} outlines a number of common analytical approaches.

The rich and high-dimensional nature of text data gives rise to a number of unique challenges for matching documents using the standard approach described above.
From a causal inference perspective, in many text corpora there is going to be substantial lack of overlap, i.e., entire types of documents in one group that simply do not exist in the other groups. 
This lack of overlap is exacerbated by the high-dimensional aspect of text: the richer the representation of text, the harder it will be to find documents similar along all available dimensions to a target document \citep{d2017overlap}. 
This makes the many design decisions required to operationalize text for matching such as defining a distance metric and implementing a matching procedure especially challenging. Distance metrics must be defined over sparse, high-dimensional representations of text in a manner that captures the subtleties of language. 
If these representations are overly flexible, standard matching procedures can fail to identify good (or any) matches in this setting due to the curse of dimensionality. 

Lack of overlap can come from substantive lack of overlap (the documents are inherently different), but also aspects of the text representation that are not substantive (this is akin to overfitting the representation model). 
Ideally a good representation and distance metric will preserve the former but not the latter.
All of the matching procedures discussed in this work can be thought of as carving out as many high quality matches as they can find, implicitly setting parts of the corpus aside to have good comparisons across groups. This is in effect \emph{isolating} \citep{zubizarreta2014isolation} a focused comparison within a larger context. In a causal context, this can shift the implied estimand of interest to only those units in the overlap region. For further discussion of the approaches commonly used to address overlap issues, see, for example, \cite{fogarty2016discrete, dehejia2002propensity, stuart2010matching}.

In addition to these difficulties, the rich nature of text data also provides an opportunity in that it lends itself to more straightforward, intuitive assessments of match quality than are typically possible with quantitative data. Specifically, while it is difficult to interpret the quality of a matched pair of units using numerical diagnostics alone due to being high dimensional, the quality of a matched pair of text documents is generally intuitive to conceptualize. With text data, human readers can quickly synthesize the vast amount of information contained within the text and quantify match quality in a way that is directly interpretable. Thus, when performing matching with text data, final match quality can be established in a manner that aligns with human judgment about document similarity. This is a version of ``thick description,'' discussed in \citet[pg. 322]{rosenbaum2010designobservational}. This also allows for comparing different matching methods to each other in order to find methods that, potentially by using more sparse representations of text or more structured distance measures, can simultaneously find more matched documents while maintaining a high degree of match quality.

\subsection{Different types of text-based confounding}

Text is quite multifaceted, but that does not necessarily mean that the researcher needs to attend to all aspects of the text in order to appropriately control for any confounding.
The confounding feature of the text may be superficial and reducible to keywords, for example whether a news story covers politics, or it may be latent and difficult to deterministically measure, like a news story's ideological content. 

In the simplest case, for example, consider a study with a single confounding feature that affects both assignment to treatment and the outcome of interest. Suppose that feature is defined as the presence in the text of a single word or phrase that is known \textit{ex ante}. Since this can be measured deterministically using the available text data, then one can easily construct a statistic to capture that confounding (e.g., a binary variable indicating whether or not each document contains the word or phrase of interest). 
In this setting, the ``best'' text matching method will be the one that produces the best balance on that single critical word or phrase, calculated directly as the difference in means between prevalence of that word or phrase in treatment corpus and its prevalence in the control corpus.

In more complex settings, it may be necessary to control for some \textit{latent} feature of the text, which might manifest in the text data as a set of related words. For instance, in the medical study described in Section~\ref{sec:med}, a patient's degree of frailty (i.e., healthiness or lack thereof) is a potentially confounding factor that is not measured numerically. 
This latent construct may manifest in the text data as a number of different key terms or phrases (e.g., ``wheelchair bound''). If all such text-based indicators for the underlying construct of interest can be identified \textit{ex ante} based on subject matter expertise and/or substantive theory, then it may be possible to directly quantify the latent variable by applying some hand-coded decision rules to the text. (In Section~\ref{sec:med}, we invert this procedure as a validation study of our more involved matching methods: if it is possible to avoid confounding by controlling for a set of pre-specified terms, then the most successful general text matching method will be the one that produces the best aggregate balance on those key words.) 
Again, in this circumstance, we may simply calculate these features for our documents and use classic matching methods from there.

The still more difficult scenario, the scenario that is the focus of this paper, is one in which the latent confounding feature of interest is challenging to measure directly, e.g., is not reducible to key words or phrases; these are the cases where we advocate for our more involved matching process that deals with general representation and distance metrics.
In particular, many studies may have important confounding features that are inherently subjective (e.g., a hospital patient's level of optimism or a news story's partisan content). 
For example, in Section~\ref{sec:media}, we control for a subjective and latent feature of news articles: the story being covered. Since there are many different stories covered across all news articles, this confounding feature is a categorical variable in high dimension. As such, while there may be keywords which perfectly identify any one story in particular, for example the flight numbers of plane crashes or the names of important figures, compiling a complete list of all such keywords would be impossible.  
It is contexts such as these that we hope matching on more general representations of text without generating a set of hand-coded and targeted covariates will still allow for principled comparisons between groups of documents.
But these automated methods may not work in a given context, and thus we also recommend in such contexts relying on human evaluation to verify that the matching process is controlling for those aspects of the text considered most critical to obtain ones ``selection on observables'' assumption.

\section{A framework for matching with text data}
\label{sec:methods}

When performing matching, different choices at each step of the process will typically interact in ways that affect both the quantity and quality of matches obtained. This can lead to different substantive inferences about the causal effects of interest. Therefore, it is important to consider the combination of choices as a whole in any application of matching. Although some guidelines and conventional wisdom have been developed to help researchers navigate these decisions, no best practices have yet been identified in general, let alone in settings with text data, where, in addition to the usual choices for matching, researchers must also consider how to operationalize the data. We extend the classic matching framework to accommodate text documents by first identifying an appropriate quantitative representation of the corpus that ideally focuses attention on those aspects we are attempting to control for, then applying the usual steps for matching using this representation. Our framework applies in settings where summary measures of text are used to define the confounding covariates, the outcomes, or both. 

The general procedure to match documents based on aspects of text that we propose is the following:

\begin{enumerate}
\item Choose a representation of the text and define explicitly the features that will be considered covariates and those, if any, that will be considered outcomes, based on this representation.\footnote{There are additional considerations and steps required when both the covariates and outcome are characterized by text; see Appendix~\ref{app:covar}.}
\item Define a distance metric to measure the similarity of two documents based on their generated covariate values that ideally focuses attention on the aspects of text considered the most important to account for (i.e., biggest potential confounders).
\item Implement a matching procedure to generate a matched sample of documents.
\item Evaluate match quality across the matched documents, and potentially repeat Steps 1-3 until consistently high quality matches are achieved.
\item Estimate the effects of interest using the final set of matched documents.
\end{enumerate}

In the subsections below, we briefly introduce a number of different choices available in steps 1-3 of the above procedure and discuss the benefits and limitations of each. These options are summarized in Table~\ref{tab:choicetab.} We then, in Section~\ref{sec:validation}, present an approach for step 4 based on a human evaluation experiment. Finally, we illustrate step 5 through two different applications in Section~\ref{sec:applied}.

\begin{table}[ht!]
	\centering
	\begin{tabular}{cccl}
		\hline
		\textbf{Step} & \textbf{Description} & \textbf{Common Choices} & \textbf{Specifications Required} \\ \hline
		\multirow{6}{*}{1} &  & Term-Document  & Dimension of vocabulary, weighting scheme,\\ 		& & Matrix & sparsity reduction\\ 
		& Text	& Statistical Topic & Dimension of vocabulary, number of topics,\\
		& representation& Model &  prior distributions, estimation procedure   \\ 
		& & Document & Embedding dimension, training data, neural\\
		& & Embedding & network architecture, estimation procedure\vspace{1mm}\\ \hline
		\multirow{4}{*}{2} &  & Exact & None \\
		& Distance & Coarsened Exact  &  Coarsening rules \\ 
		& metric	& Continuous  &  Functional form (e.g., Euclidean, Cosine,\\
		& & & Mahalanobis, propensity score) \\ \hline
		3 & Matching & Nearest neighbor & Replacement, caliper, trimming \\
		&										procedure & Optimal  & Caliper, trimming objective function \\
		&								& Cardinality & Caliper, trimming \\ \hline
		
	\end{tabular}
	\caption{Common choices at each of the first three stages of the text matching procedure and examples of additional specifications required by each choice.}
	\label{tab:choicetab}
\end{table}

These steps and choices required to perform matching should be familiar to those with experience in standard matching, as many of the choices are directly parallel to a standard matching procedure. 
Because text is such a rich source of data, however, how our decisions connect to which aspects of our data are most important to match on is much more salient as compared to classic matching. 
The choices made in the above steps boil down to making an expert judgment as to what aspects of the text give one the best approximation of selection on observables. 
In our media example, to illustrate, we judged that matching on the story covered was the most important thing and therefore generated a human evaluation experiment that targeted this particular measure and selected the procedure that performed this task the best.
For a more thorough discussion and description of the various choices within these steps, see Appendix~\ref{app:text}.

\subsection{Text representations} 

The \textit{representation} of a text document transforms an ordered list of words and punctuation into a vector of covariates, and is the most novel necessary component of matching with text. 
To choose a representation, the researcher must first formulate a definition for textual similarity that is appropriate for the study at hand. 
In some cases, all of the information about potential confounders captured within the text data may be either directly estimable (e.g., frequency of a particular keyword) or may be plausible to estimate using a single numerical summary (e.g., the primary topic of a document estimated using a topic model).
In other cases, such a direct approach may not be possible.

The most common general representation of text is as a ``bag-of-words,'' containing unigrams and often bigrams, collated into a term-document matrix (TDM); the TDM may also be rescaled according to Term Frequency-Inverse Document Frequency (TF-IDF) weighting. Without additional processing, however, these vectors are typically very long; more parsimonious representations involve calculating a document's factor loadings from unsupervised learning methods like factor analysis or Structural Topic Models (STM) \citep{roberts2016model}, or calculating a scalar propensity score for each document using the bag-of-words representation \citep{taddy2013multinomial}. Finally, we also consider a Word2Vec representation \citep{mikolov2013distributed}, in which a neural network embeds words in a lower-dimensional space and a document's value is the weighted average of its words. 

Each of these methods involves a number of tuning parameters. When using the bag-of-words representation, researchers often remove very common and very rare words at arbitrary thresholds, as these add little predictive power, or choose to weight terms by their inverse document frequency; these pre-processing decisions can be very important \citep{denny2018text}. Topic models such as the STM are similarly sensitive to these pre-processing decisions \citep{fan2017promoting} and also require specification of the number of topics and selecting covariates, which are often unstable. Word2vec values depend on the dimensionality of the word vectors as well as the training data and the architecture of the neural network.

Overall, when choosing a representation, researchers need to consider what aspects of the text are confounding the outcome. 
For example, in our evaluation study that used matched pairs of news articles from Fox News and CNN, we were interested in identifying pairs of stories that were about the same general topic (e.g., plane crashes versus public policy) and that also utilized the same set of keywords (e.g., ``AirAsia'' or ``Obama''); this may suggest (as we found) that representations that preserve the details of different keywords was important for obtaining good matches. 
Generally, when the objective is to identify exact or nearly exact matches, we recommend using text representations that retain as much information in the text as possible. 
In particular, documents that are matched using the entire term-vector will typically be similar with regards to both topical content and usage of keywords, while documents matched using topic proportions may only be topically similar. 

When the aspects of text are more targeted or specific, simply directly computing the relevant covariates constructed by hand-coded rules may be the best option.
That being said, one might imagine that generally matching on the content of the text---as represented by the specific words and phrases used---will frequently capture much of what different researchers in different contexts may view as the necessary component for their selection on observables assumption. 
Clearly this is an area for future work; as we see more matching with text in the social sciences, we will also see a clear picture as to what structural aspects of text are connected to the substantive aspects of text that researchers find important.

\subsection{Distance metrics} Having converted the corpus into covariate representations, the second challenge is in \textit{comparing} any two documents under the chosen representation to produce a measure of distance. The two main categories of distance metrics are exact (or coarsened exact) distances, and continuous distances. Exact distances consider whether or not the documents are identical in their representation. If so, the documents are a match. Coarsened exact distance bins each variable in the representation, then identifies pairs of documents which share the same bins. If the representation in question is based on a TDM, these methods are likely to find only a small number of high quality matches, given the large number of covariates that all need to agree either exactly or within a bin. The alternative to exact distance metrics is continuous distance metrics such as Euclidean distance, Mahalanobis distance, and cosine distance. Counter to exact and coarsened exact metrics, which identify matches directly, these metrics produce scalar values capturing the similarity between two documents. 

\subsection{Matching procedures} After choosing a representation and a distance metric, the choice of matching procedure often follows naturally, as is the case in standard matching analyses. Exact and coarsened exact distance metrics provide their own matching procedure, while continuous distance metrics require both a distance formula and a \textit{caliper} for specifying the maximum allowable distance at which two documents may be said to still match. The calipers may be at odds with the desired number of matches, as some treated units may have no control units within the chosen caliper, and may subsequently be ``pruned'' by many common matching procedures. Alternatively, researchers may allow any one treated unit to match multiple controls, or may choose a greedy matching algorithm.

\section{Experimental evaluation of text matching methods}
\label{sec:validation}
In the previous section, we presented different forms of representations for text data and described a number of different metrics for defining distance using each type of representation. Any combination of these options could be used to perform matching. However, the quantity and quality of matches obtained depend heavily on the chosen representation and distance metric. For example, using a small caliper might lead to only a small number of nearly-exact matches, while a larger caliper might identify more matches at the expense of overall match quality. Alternatively, if CEM on a STM-based representation produces a large number of low-quality matches, applying the same procedure on a TDM-based representation may produce a smaller number of matches with more apparent similarities. 

We investigate how this quantity versus quality trade-off manifests across different combinations of methods through an evaluation experiment performed with human subjects. Applying several variants of the matching procedure described in Section \ref{sec:methods} to a common corpus, we explore how the quantity of matched pairs produced varies with different specifications of the representation and distance metric. Then, to evaluate how these choices affect the quality of matched pairs, we rely on evaluations of human coders. 

In this study, we consider five distance metrics (Euclidean distance, Mahalanobis distance, cosine distance, distance in estimated propensity score, and coarsened exact distance), as well as 26 unique representations,\footnote{Because estimation and distance calculations with high-dimensional text representations can be computationally intensive, we restrict our analyses to this set of 26 possible representations, which we believe provide an adequate representation of the spectrum of possible text-representations that could be used for applications of text-matching. However, we emphasize that the methods presented in this paper, including the procedure for text-matching and the framework for performing systematic evaluations of text-matching methods, can be extended to include any number of additional variants to the representations considered here.} including nine different TDM-based representations, 12 different STM-based representations, and five Word2Vec embedding-based representations. Crossing these two factors produces 130 combinations, where each combination corresponds to a unique specification of the matching procedure described in Section \ref{sec:methods}. Among these combinations, 5 specifications are variants of the TIRM procedure developed in \cite{roberts2018adjusting}. Specifications of each of the procedures are provided in Appendix \ref{app:index}.

To compare the different choices of representation and distance metric considered here, we apply each combination to a common corpus to produce a set of matched pairs for each.
We use a corpus of $N=3,361$ news articles published from January 20, 2014 to May 9, 2015, representing the daily front matter content for each of two online news sources: Fox News ($N=1,796$) and CNN ($N=1,565$). The news source labels were used as the treatment indicator, with $Z=1$ for articles published by Fox News and $Z=0$ for articles published by CNN. 
To match, we first calculate the distances between all possible pairs of treated and control units based on the specified representation and distance metric. Each treated unit is then matched to a set of control units with whom its distance was within the specified caliper.\footnote{For each of the combinations that did not use the CEM metric, the caliper was calculated as the 0.1th quantile of the distribution of distances under that combination for all 1796 $\times$ 1565 = 2,810,740 possible pairs of articles.} Using this procedure, 13 of the original 130 specifications considered did not identify any matched pairs. The union of matched pairs identified across the remaining 117 procedures resulted in 30,647 unique pairs.

Each procedure identified between 41 and 1605 total pairs of matched articles, with an average of 502 pairs produced per matching procedure. These pairs covered between 69 to 2942 unique articles within the corpus. Specifically, each procedure identified one or more matches for between 34 (2\%) and 1566 (87\%) of the 1796 unique articles published by Fox News and identified matches for between 20 (1\%) and 1376 (88\%) of the 1565 unique CNN articles. 

Conversely, each of the 30,647 unique pairs of matched articles was identified, on average, by 1.91 of the 117 different procedures, with 6,910 (22.5\%) of unique pairs matched by between 2 to 55 of the 117 procedures and the remaining 23,737 pairs matched by only one procedure. We view the frequency of each unique pair within the sample of 58,737 pairs identified as a rough proxy for match quality because, ideally when performing matching, the final sample of matched pairs identified will be robust to different choices of the distance metric or representation. Thus, we expect that matched pairs that are identified by multiple procedures will have higher subjective match quality than singleton pairs.

\subsection{Measuring match quality}
\label{ssec:matchqual}
In standard applications of matching, if two units that are matched do not appear substantively similar, then any observed differences in outcomes may be due to poor match quality rather than the effect of treatment. Usual best practice is to calculate overall balance between the treatment and control groups, which is typically measured by the difference-in-means for all covariates of interest. If differences on all matched covariates are small in magnitude, then the samples are considered balanced, and thus, typically, well-matched.

As previously discussed, to calculate balance in settings where the covariates are text data, these standard balance measures typically fail to capture meaningful differences in the text. Further, due to the curse of dimensionality in these settings, it is likely that at least some (and probably many) covariates will be unbalanced between treatment and control groups. Thus, to measure match quality we rely on a useful property of text: its ease of interpretability. A researcher evaluating two units that have been matched on demographic covariates, for example, may be unable to verify the quality of a matched pair. 
However, depending on what aspects of text the researcher is substantively attempting to match on, human coders who are tasked with reading two matched text documents are often amply capable of quantifying their subjective similarity if given instructions as to what to attend to. 
We leverage this property to measure match quality using an online survey of human respondents, where match quality is defined on a scale of 0 (lowest quality) to 10 (highest quality). 

To obtain match quality ratings, we conducted a survey experiment using Amazon's Mechanical Turk (MTurk) and the Digital Laboratory for the Social Sciences (DLABSS) \citep{dlabss}. Online crowd-sourcing platforms such as these have been shown to be effective for similarity evaluations in a number of settings \citep{mason2012conducting}. For instance, a study by \cite{snow2008cheap} that tasked non-expert human workers on MTurk with five natural language evaluations reported a high degree of agreement between the crowd-sourced results and gold-standard results provided by experts. In the present study, respondents were first informed about the nature of the task and then given training\footnote{For training, participants were first informed about the nature of the task. Next, participants were presented with a scoring rubric and were informed to use this rubric as ``a guide to help [them] determine the similarity of a pair of articles.'' In the final component of training, participants were asked to read and score three pre-selected pairs of articles, which were chosen to represent pairings that we believe have match quality scores of zero, five, and ten, respectively. After scoring each training pair, participants were informed about the anticipated score for that pair and provided with an explanation for how that determination was made.} on how to evaluate the similarity of two documents. After completing training, participants were then presented with a series of 11 paired newspaper articles, including an attention check and an anchoring question, and asked to assign a similarity rating. For each question, participants were instructed to read both articles in the pair and rate the articles' similarity from zero to ten, where zero indicates that the articles are entirely unrelated and ten indicates that the articles are covering the exact same event. Snapshots of the survey are presented in Appendix \ref{app:surveysnap}. 

We might be concerned that an online convenience sample may not be an ideal population for conducting this analysis, and that their perceptions of article similarity might differ from the overall population, or from trained experts. To assess the reliability of this survey as an instrument for measuring document similarity, we leverage the fact that we performed two identical pilot surveys prior to the experiment using respondents from two distinct populations and found a high correlation ($\rho=0.85$) between the average match quality scores obtained from each sample. Additional details about this assessment are provided in Appendix \ref{app:reliability}. We take note that these populations, MTurkers and DLABSS respondents, are both regularly used as coders to build training data sets for certain tasks in machine learning; the hallmark of these tasks is that they are easily and accurately performed by untrained human respondents. We argue that this task of identifying whether two articles discuss related stories falls squarely in this category, and our inter-coder reliability test in Appendix \ref{app:reliability} supports this argument.\footnote{For researchers interested in conducting their own text matching evaluation studies, we note that MTurk and DLABSS populations may not always be applicable, especially in contexts where domain expertise is required. }

In an ideal setting, for each unique matched pair identified using the procedure described above, we would obtain a sample of similarity ratings from multiple human coders. Aggregating these ratings across all pairs in a particular matched data set would then allow us to estimate the average match quality corresponding to each of the 130 procedures considered, with the quality scores for the 13 procedures that identified no matches set to zero. Though this is possible in principle, to generate a single rating for each unique matched pair requires that a human coder read both documents and evaluate the overall similarity of the two articles. This can be an expensive and time-consuming task. Thus, in this study, it was not possible to obtain a sample of ratings for each of the 30,647 unique pairs. 

Instead, we took a stratified, weighted sample of pairs such that the resulting sample would be representative of the population of all 30,647 unique matched pairs as well as the population of 2,780,093 pairs of documents that were not identified by any of the matching procedures. Specifically, the sample was chosen such that each of the 130 matching procedures that identified a non-zero number of matches would be represented by at least four pairs in the experiment. For each stratum, the sampling weights for each pair were calculated proportional to the estimated match quality of that pair, calculated using a predictive model trained on human-coded data from a pilot experiment. We also sampled an additional 50 unique pairs from the pool of 2,780,093 pairs not identified by any matching procedures. 
Ratings obtained from these pairs can be used to obtain a reference point for interpreting match quality scores. The resulting sample consisted of 505 unique pairs ranging the full spectrum of predicted match quality scores. Each respondent's set of nine randomly selected questions were drawn independently such that each pair would be evaluated by multiple respondents. Using this scheme, each of the 505 sampled pairs was evaluated by between six and eleven different participants (average of 9). Question order was randomized, but the anchor was always the first question, and the attention check was always the fifth question. 

We surveyed a total of 505 respondents. After removing responses from 52 participants who failed the attention check,\footnote{The attention check consisted of two articles with very similar headlines but completely different article text. The text of one article stated that this question was an attention check, and that the respondent should choose a score of zero. Participants who did not assign a score of zero on this question are regarded as having failed the attention check.} all remaining ratings were used to calculate the average match quality for each of the 505 sampled pairs evaluated.
These scores were then used to evaluate each of the 130 combinations of methods considered in the evaluation, where the contribution of each sampled pair to the overall measure of quality for a particular combination of methods was weighted according to its sampling weight. This inferential procedure is described more formally in Appendix \ref{app:math}.

\subsection{Results}
\subsubsection{Which automated measures are most predictive of human judgment about match quality?} 

Our primary research question concerns how unique combinations of text representation and distance metric contribute to the quantity and quality of obtained matches in the interest of identifying an optimal combination of these choices in a given setting. We can estimate the quality of the 130 matching methods considered in the evaluation experiment using weighted averages of the scores across the 505 pairs evaluated by human coders. However, it is also of general interest to be able to evaluate new matching procedures without requiring additional human experimentation. We also want to maximize the precision of our quality estimates for the 130 methods considered in this study. To these ends, we examine if we can predict human judgment about match quality based on the distance scores generated by each different combination of one representation and one distance metric. If the relationship between the calculated match distance and validated match quality is strong, then we may be confident that closely-matched documents, as rated under that metric, would pass a human-subjects validation study.

\begin{figure}
     \begin{center}
\includegraphics[width=0.9\textwidth]{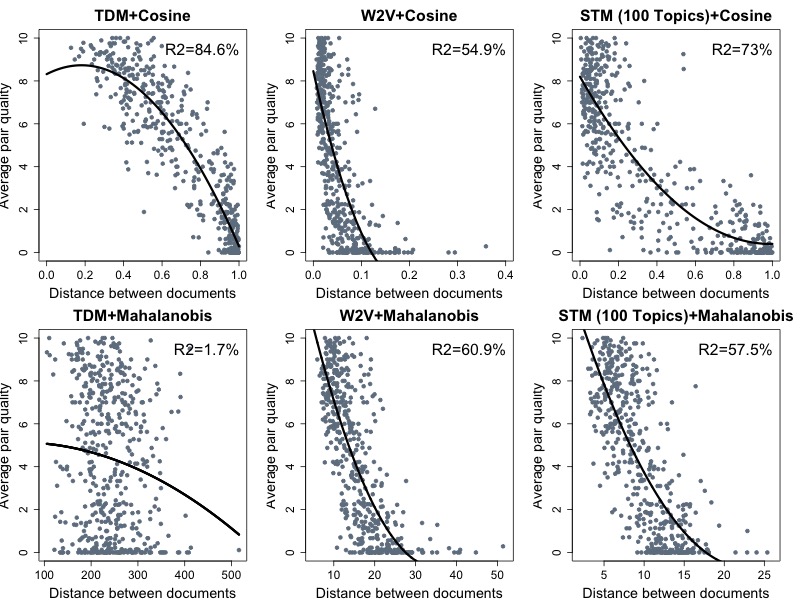}
	\end{center}
	 \caption{Distance between documents and match quality based on the cosine distance measured over a TDM-based representation (top left) exhibit a stronger relationship than cosine distance measured over both a W2V-based representation (top center) and a STM-based representation (top right), and a much stronger relationship than the Mahalanobis distance measured over a TDM-based representation (bottom left), a W2V-based representation (bottom center) or a STM-based representation (bottom right).}
     \label{fig:scatter}
\end{figure}

To evaluate the influence of each distance score on match quality, we take the pairwise distances between documents for each of the 505 matched pairs used in the evaluation experiment under different combinations of the representations and distance metrics described in Section~\ref{sec:methods}. After excluding all CEM-based matching procedures, under which all pairwise distances are equal to zero or infinity by construction, all distances were combined into a data set containing 104 distance values for each of the 505 matched pairs. Figure~\ref{fig:scatter} gives six examples of how these distances correlate with observed match quality based on human ratings of similarity, along with the fitted regression line obtained from quadratic regressions of average match quality on distance. Here, the strong correlations suggest that automated measures of match quality could be useful for predicting human judgment.
The particularly strong relationship between the cosine distance metric calculated over a TDM-based representation provides additional evidence in favor of matching using this particular combination of methods. These findings also suggest that the increased efficiency achieved with TDM cosine matching is not attributable to the cosine distance metric alone, since the predictive power achieved using cosine distance on a Word2Vec (W2V) representation or a STM-based representation is considerably lower than that based on a TDM-based representation.

\begin{figure}
     \begin{center}
\includegraphics[width=0.5\textwidth]{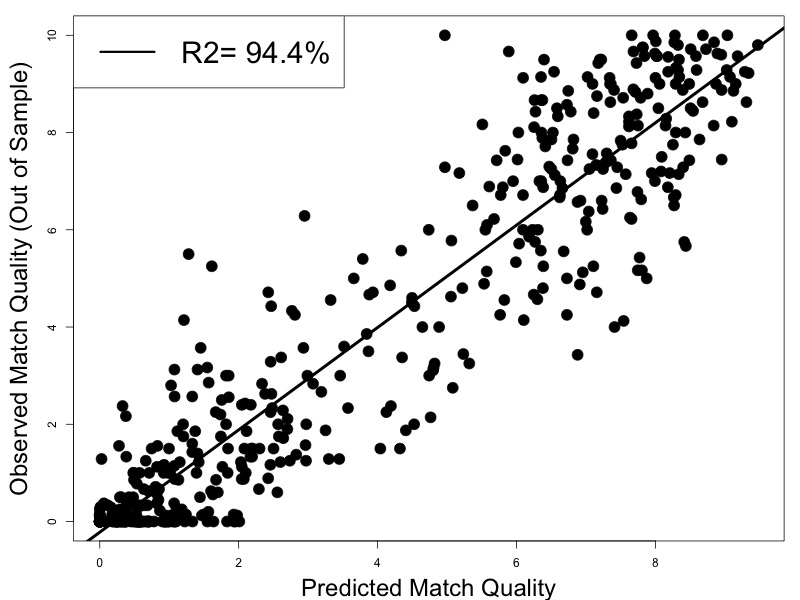}
	\end{center}
\caption{Predictive model for match quality trained on human evaluations has a correlation of 0.944 with observed quality scores obtained in a separate human evaluation experiment on a different set of pairs, indicating high out-of-sample predictive accuracy.}
     \label{fig:lasso}
\end{figure}

To leverage the aggregate relationship of the various machine measures of similarity on match quality, we developed a model for predicting the quality of a matched pair of documents based on the 104 distance scores, which we then trained on the 505 pairs evaluated in our survey experiment. For estimation, we use the LASSO \citep{tibshirani1996regression}, implemented with ten-fold cross validation \citep{kohavi1995study}. Here, for each of the 505 pairs, the outcome was defined as the average of the ratings received for that pair across the human coders, and the covariates were the 104 distance measures. We also included quadratic terms in the model, resulting in a total of $p$=208 terms. Of these, the final model obtained from cross-validation selected 19 terms with non-zero coefficients. 
However, our results suggest that the majority of the predictive power of this model primarily comes from two terms: cosine distance over the full, unweighted term-document matrix and cosine distance over an STM with 100 topics. Figure \ref{fig:lasso} shows the out-of-sample predictive performance of the model for a distinct sample of 472 pairs of documents evaluated in a separate survey experiment.
The correlation of predictions to measured quality for this sample was approximately 94\%. In sample correlation was 89\% (the stronger out-of-sample correlation is likely driven by a different distribution of matched pairs evaluated).To evaluate the sensitivity of this model to the chosen regularization scheme, we performed a similar analysis using ridge regression and found only a negligible difference in predictive performance.

The high predictive accuracy of our fitted model suggests that automated measures of similarity could be effectively used to evaluate new matched samples or entirely new matching procedures without requiring any additional human evaluation.\footnote{Since this model was trained on human evaluations of matched newspaper articles, extrapolating predictions may only be appropriate in settings with similar types of documents. However, our experimental framework for measuring match quality could be implemented using text data to build a similar predictive model in other contexts.} We can also use it to enhance the precision of our estimates of match quality for the 130 matching methods considered in the evaluation experiment using model-assisted survey sampling  methods.

\subsubsection{Which methods make the best matching procedures?}

\begin{figure}[h]
     \begin{center}
\includegraphics[width=0.9\textwidth]{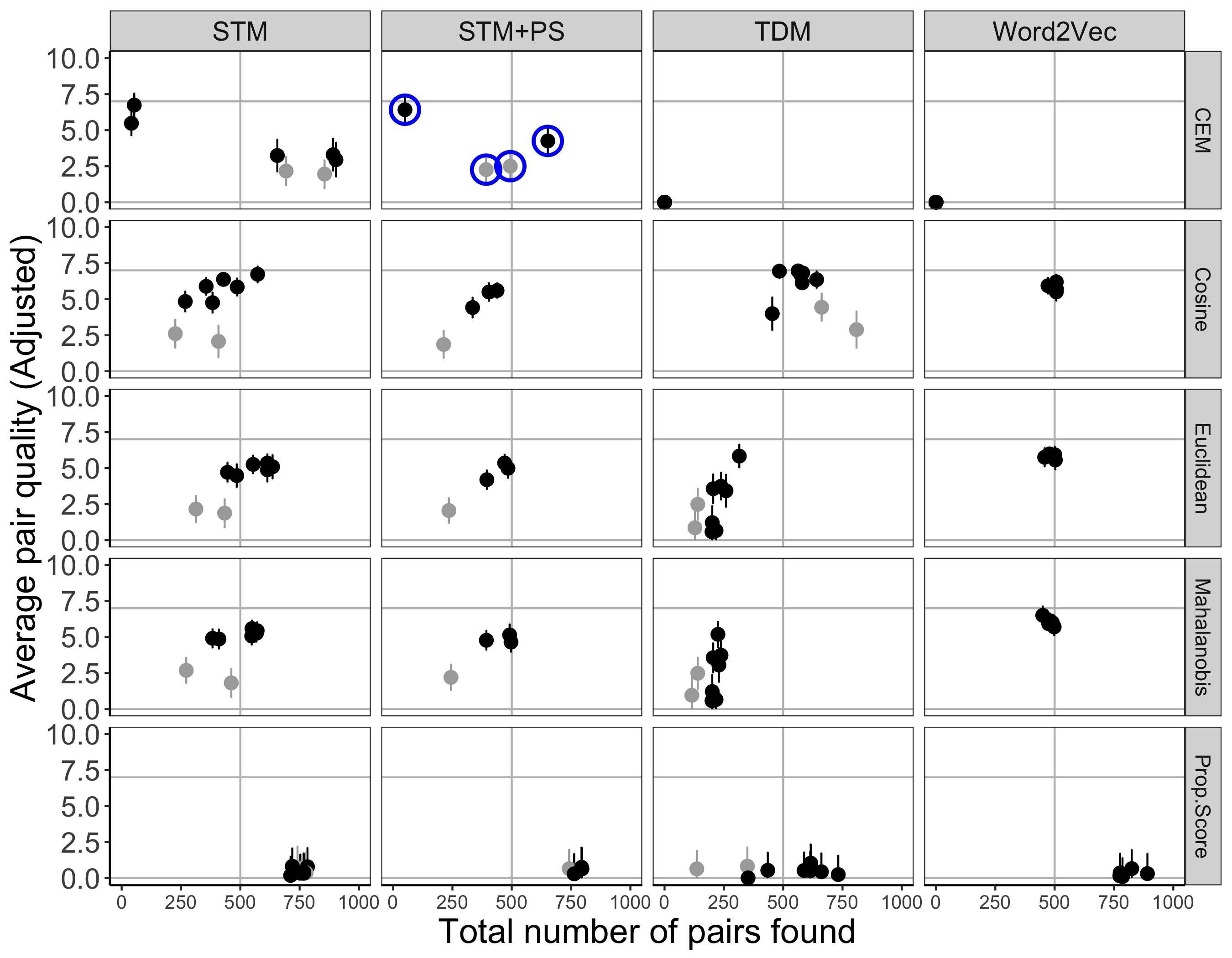}
	\end{center}
	 \caption{Number of matches found versus average model-assisted match quality scores for each combination of matching methods. Grey points indicate procedures with extreme reduction in information (e.g., procedures that match on only stop words). Blue circles highlight procedures that use existing state-of-the-art methods for text matching. One procedure with many low quality pairs at coordinates (1605,1.39) is excluded from this plot.}
     \label{fig:qualityAll}
\end{figure}

To compare the performance of the final set of 130 matching procedures considered in our study, we, for each method, estimate the average quality of all pairs selected by that method.
We increase precision of these estimates using model-assisted survey sampling.
In particular, we first use the predictive model described above to predict the quality of all matched pairs of a method.
This average quality estimate is then adjusted by a weighted average of the residual differences between predicted and actual measured quality for those pairs directly evaluated in the human experiment.
(The average quality scores for the 13 procedures that identified no matches are all set equal to zero.)
This two-step process does not depend on the model validity and is unbiased.\footnote{Nearly unbiased that is. There is a small bias term due using a Ha\"jek-style approach rather than Horvitz-Thompson. This comes from the sample having a random total weight due to using the weighted sampling method.} 
We assess uncertainty with a variant of the parametric bootstrap.
See Appendix~\ref{app:math} for further details of the estimation approach and associated uncertainty quantification.
Figure~\ref{fig:qualityAll} shows the performance of each of the 130 procedures in terms of average predicted match quality vs. number of pairs identified, with uncertainty intervals estimated using a parametric bootstrap.
We group the procedures by the large-scale choices of representation and distance metric used.
Within each tile of the larger plot are different procedures corresponding to different design decisions within a general approach such as tuning parameters such as number of topics used in a topic model.
As sensitivity check, see Appendix~\ref{app:raw_results} for results using the simple weighted means of the sampled pairs of each method; results are broadly similar.

Table~\ref{tab:qualitytab} summarizes the performance of the different methods considered grouped by representation. The methods which generally produce the highest quality matches for our study are those based on cosine distance calculated over a TDM-based representation. The method that produces the most matches out of all 130 procedures considered uses STM on ten topics with sufficient reduction and CEM in 2 bins and identifies 1605 matched pairs. However, this method is among the lowest scoring methods in terms of quality, with a sample-adjusted average match quality of 1.41. Conversely, a procedure that uses STM on 30 topics with sufficient reduction and CEM in 3 bins, appears to produce considerably higher quality matches, with an average match quality of 5.69, but identifies only 50 matched pairs. In comparison, a method that combines a bounded TDM with TF-IDF weighting with the cosine distance metric identified 582 matches with an average match quality of 7.06. This illustrates an important weakness of CEM: too few bins produce many low quality matches, while too many bins produce too few matches, even though they are high quality. While in many applications there may be a number of bins which produce a reasonable number of good quality matches, that is not the case in our setting. Here, two bins produce poor matches while three bins produce far too few. This tradeoff does not appear to be present for matching procedures using cosine distance with a TDM-based representation, which dominate in both number of matches found and overall quality of those matched pairs. In addition, the matching procedures based on this combination appear to be more robust to various the pre-processing decisions made when constructing the representation than procedures that use an alternative distance metric or representation, as illustrated by the tight clustering of the variants of this general approach on the plot.

\begin{table}[htbp]
	\centering
	\caption{Ranges for the number of pairs identified, observed procedure-level quality, predicted procedure-level quality, and corresponding standard errors for both quality measures. The second column indicates the distance metric that achieved the highest average predicted match quality when used in combination with each type of representation. }
	\begin{tabular}{llccccc}
		\hline
     \multirow{2}{*}{Representation}     &     Preferred  &   Number of    & \multicolumn{2}{c}{Average Match Quality} & \multicolumn{2}{c}{Standard Error}  \\ \cline{4-5}\cline{6-7}
 & Metric & Pairs & Unadjusted & Adjusted & Unadjusted & Adjusted \\ \hline
STM   & Cosine & 41-1893 & 0.11 - 6.67 &0.41 - 6.35 & 0.24 - 0.76 & 0.48 - 1.40 \\
STM+PS & CEM   & 50-1605 & 0.30 - 6.97 & 0.54 - 5.70 & 0.28 - 0.75 & 0.59 - 1.34 \\
TDM   & Cosine & 115-809 & 0.11 - 7.42 & 0.37 - 7.16 & 0.21 - 0.69 & 0.41 - 1.33 \\
Word2Vec & Cosine & 450-891 & 0.05 - 7.85 & 0.41 - 6.37 & 0.25 - 0.70 & 0.48 - 1.34 \\ \hline
\end{tabular}%
	\label{tab:qualitytab}%
\end{table}%

Overall, our results indicate that, in our context, matching on the full TDM produces both more and higher quality matches than matching on a vector of STM loadings when considering the content similarity of pairs of news articles. Moreover, TDM-based representations with cosine matching appear relatively robust to tuning parameters including the degree of bounding applied and the choice of weighting scheme. STM-based representations, on the other hand, appear to be somewhat sensitive to tuning parameters, with representations that include a large number of topics achieving higher average match quality than those constructed over a smaller number of topics. This result provides further support for the findings in \cite{roberts2018adjusting}. In that paper, the authors found that matching on more topics generally led to better results in terms of recovering pairs of nearly identical documents.

\subsection{Evaluating text matching methods}

In our applied examples, we find that text representations that use the TDM or Word2Vec embeddings paired with cosine distance achieve the best results in terms of maximizing predicted match quality and the number of matches identified.
But these results may well not be general.
We therefore emphasize that applied researchers conducting their own text matching analyses need to conduct their own systematic evaluations to determine which representations and distance metrics work best in their domains. 
Here we offer some thoughts on how to think about and design convenient and flexible systematic evaluations.

First, until we have more general research knowledge in the field, we recommend implementing a suite of text matching procedures that include a diverse set of representations and distance metrics, and then comparing the matches identified by the different methods. 
If there is substantial overlap across all methods, it may be that no evaluation is necessary. 
However, in most cases, the sets of identified matches will largely diverge.
In this case, we recommend formally evaluating which methods best capture the confounding previously identified.

In our discussion of different types of confounding, above, we noted that if the aspects of text that are most important are directly measurable, then these more general text matching approaches are not needed, strictly speaking.
In this case we recommend directly assessing balance on specific covariates built from the text.
But if general text matching methods are used in such contexts, we believe checking these core covariates to still be of use as signals as to which methods are at least achieving balance on some core summary statistics.
This is akin to viewing mean balance as a proxy for covariate balance in classic matching.

If the potential confounding truly hinges on the more complex and latent aspects of text, however, then one could ideally leverage human judgment to hand evaluate the full set of possible matched pairs of text documents.
In our case, for example, we could, given unlimited resources, ask human coders to read through the entire corpus of news articles and put them into bins according to which stories they cover. Even untrained human coders could be reliably good at this task.
This, of course, is generally not possible, but we hope the methods described above serve a similar function.

As we have seen, we can evaluate the success of such an attempt by inverting the full human-coding procedure to generate a test: we identify a set of possible matches using automated text matching methods and then and present a subset of them to trained human coders. These human coders can then evaluate sample pairs of matched documents to determine which matches are systematically ``best'' according to their own judgment. 
Using this information we can then see which methods appear to best match on the targeted aspects of text.
This human coding task is of utmost importance, requiring both careful pretesting and substantial guidance to ensure the humans attend to the aspects of text deemed most important as potential confounders. 
In particular, the primary concern is instructing the human coders to evaluate similarity \textit{along the latent dimension of interest}, which in our media case is whether any two articles truly cover the same events or issues.

One final circumstance bears discussion: it may be the case that the identified latent dimension of interest is challenging or impossible for human evaluators to reliably code. For example, even two experienced medical doctors may systematically disagree in their readings of patient data such as X-rays \citep{steiner2018impact}. In such cases, human evaluations may not serve as a reliable ground truth to which automated text match quality may be compared. It is still possible that automated text matching methods would work well in these cases, but researchers cannot validate those results in this framework. 

This and other open questions, including identifying what contexts would have topic model- or propensity score-based representations outperform TDM-based or Word2Vec embedding-based representations, we leave to future research.

\section{Applications}
\label{sec:applied}

\subsection{Decomposing media bias}\label{sec:media}
While American pundits and political figures continue to accuse major media organizations of ``liberal bias,'' scholars, after nearly two decades of research on the issue, have yet to come to a consensus about how to \textit{measure} bias, let alone determine its direction. A fundamental challenge in this domain is how to disentangle the component of bias relating to \emph{how} a story is covered, often referred to as ``presentation bias'' \citep{groseclose2005measure, gentzkow2006media,ho2008measuring,gentzkow2010drives, groeling2013media}, from the component relating to \emph{what} is covered, also known as ``selection bias'' \citep{groeling2013media} or ``topic selection.''
In particular, systematic comparisons of \emph{how} stories are covered by different news sources (e.g., comparing the level of positive sentiment expressed in the article) may be biased by differences in the content being compared. We present a new approach for addressing this issue by using text matching to control for selection bias. 

We analyze a corpus consisting of $N=9,905$ articles published during 2013 by each of 13\footnote{The original data included 15 news sources, but BBC and The Chicago Tribune are excluded from this analysis due to insufficient sample sizes for these sources} popular online news outlets. This data was collected and analyzed in \cite{budak2016fair}. The news sources analyzed here consist of Breitbart, CNN, Daily Kos, Fox News, Huffington Post, The Los Angeles TImes, NBC News, The New York Times, Reuters, USA Today, The Wall Street Journal, The Washington Post, and Yahoo. In addition to the text of each article, the data include labels indicating each articles' primary and secondary topics, where these topics were chosen from a set of 15 possible topics by human coders in a separate evaluation experiment performed by \cite{budak2016fair}. 
The data also include two human-coded outcomes that measure the ideological position of each article on a 5-point Likert scale. Specifically, human workers tasked with reading and evaluating the articles were asked ``on a scale of 1-5, how much does this article favor the Republican party?'', and similarly,  ``on a scale of 1-5, how much does this article favor the Democratic party?'' 

To perform matching on this data, we use the optimal procedure for identifying articles covering the same underlying story identified by our prior evaluation experiment: cosine matching on a bounded TDM.\footnote{Since the outcomes of interest in this analysis are human-coded measures of favorability toward democrats and republicans, we limit the vocabulary of the TDM to include only nouns and verbs to avoid matching on aspects of language that may be highly correlated with these outcomes.}  
Because in this example we have a multi-valued treatment with 13 levels, each representing a different news source, we follow the procedure for template matching\footnote{To implement the template matching procedure, we first generate a template sample of $N=150$ articles chosen to be the most representative of the corpus in terms of the distribution of primary topics among 500 candidate samples of this size. Once this template is chosen, for each treatment level (i.e., news source), we then perform optimal pair matching within primary topics to identify a sample of 150 articles from that source that most closely match the template sample with regards to cosine distance calculated over the TDM. 
Iterating through each of the 13 target sources, this produces a final matched sample of $13 \times 150 = 1,950$ matched articles.} described in \cite{silber2014template} to obtain matched samples of 150 articles across all treatment groups. In brief, the template matching procedure first finds a representative set of stories across the entire corpus, and uses that template to find a sample of similar articles within each source that collectively cover this canonical set of topics. This allows us to identify a set of articles sampled from each source that are all similar to the same template and therefore similar to each other.

Before matching, our estimates of a news source's average favorability are a measure of overall bias, which includes biases imposed through differential selection of content to publish as well as biases imposed through the language and specific terms used when covering the same content. 
The matching controls selection biases due to some sources selecting different stories that may be more or less favorable to a given party than other stories.
Differences in estimated favorability on the matched articles can be attributed to presentation bias. 
The difference between estimates of average favorability before matching (overall bias) and estimates after matching (presentation bias) therefore represent the magnitude of selection biases imposed by the sources. 
Large differences between pre- and post-matched estimates indicate a stronger influence of selection bias relative to presentation bias.

\begin{figure}
     \begin{center}
     \includegraphics[width=0.9\textwidth]{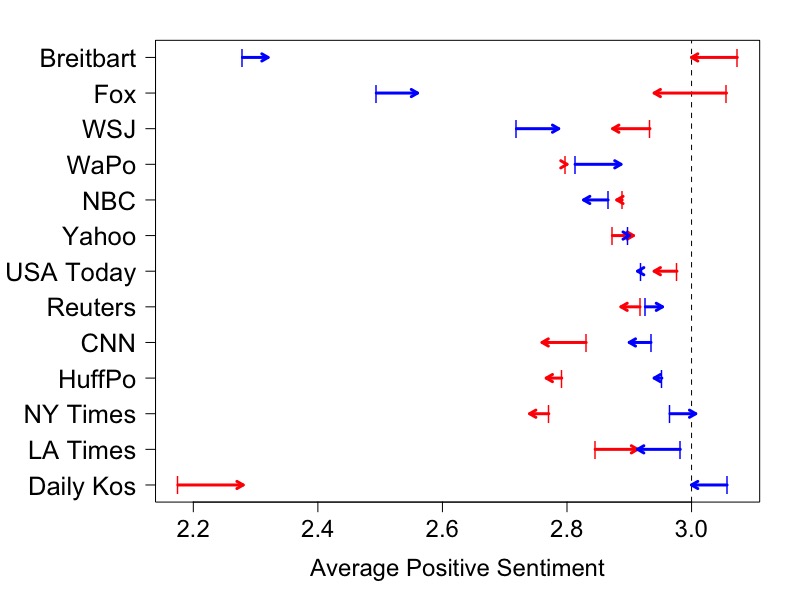}
	\end{center}
	 \caption{Estimates of average favorability toward Democrats (blue) and Republicans (red) for each source both before and after matching.}
     \label{fig:allmeans}
\end{figure}

Figure~\ref{fig:allmeans} shows the average favorability toward Democrats (blue) and Republicans (red) for each news source overall, and the average favorability among the template matched documents. 
Arrows begin at the average score before matching, and terminate at the average score after matching. 
The length of the arrows is the estimated magnitude of the bias of each source that is attributable to differences in selection. 

Before discussing the pattern of shifts, we first look at overall trends of favorability across sources.
First, overall sentiment towards Republicans generally hovers around 2.8 to 3.1, slightly less, on average, than the partisan neutrality of $x=3$, which corresponds to a response of ``neither favorable nor unfavorable.''
The one exception is the Daily Kos, which is unfavorable.
Other sources (CNN, the Huffington Post, the NY Times, and the LA Times) are at the low end of this range, indicating some negative sentiment.
For the Democrats, there is somewhat more variation, however, with Brietbart being the least favorable, followed by Fox and WSJ, and the Daily Kos being the most.

Furthermore, it is primarily the more extreme sources that show selection effects.
Breitbart, Fox and WSJ, for example, all become more positive towards Democrats and less positive towards Republicans when we adjust for story.
This suggests they tend to select stories that are biased more towards Republicans and away from Democrats, a selection bias effect.
Similarly, the LA Times and Daily Kos show the opposite trends, again showing selection bias effects in the opposite direction.
The remaining sources do not appear significantly be impacted by controlling for selection.

We performed a series of sensitivity checks to assess the stability of our results to different specifications of the matching procedure and/or different choices of template sample.
We also examine the variability due to randomly matching documents to assess how much estimation uncertainty is present in our analysis.
Details of these analyses are provided in Appendix ~\ref{app:sensitivity_media}.
Generally, we see that estimating the selection effect of an individual source is difficult, and that the magnitude of the selection effects tends to be small, indicating that the choice of what stories to cover is not driving the overall favorability ratings.
In other words, most differences in favorability appear to be driven by presentation bias.

\subsection{Improving covariate balance in observational studies}\label{sec:med}
In our second application, we demonstrate how text matching can be used to strengthen inferences in observational studies with text data. Specifically, we show that text matching can be used to control for confounders measured by features of the text that would otherwise be missed using traditional matching schemes.

We use a subset of the data first presented in \cite{feng2018tte}, which conducted an observational study designed to investigate the causal impact of bedside transthoracic echocardiography (TTE), a tool used to create pictures of the heart, on the outcomes of adult patients in critical care who are diagnosed with sepsis. The data were obtained from the Medical Information Mart for Intensive Care (MIMIC) database \citep{johnson2016mimic} on 2,401 patients diagnosed with sepsis in the medical and surgical intensive care units at a Massachusetts Institute of Technology university hospital located in Boston, Massachusetts. Within this sample, the treatment group consists of 1,228 patients who received a TTE during their stay in the ICU (defined by time stamps corresponding to times of admission and  discharge) and the control group is comprised of 1,173 patients who did not receive a TTE during this time. For each patient we observe a vector of pre-treatment covariates including demographic data, lab measurements, and other clinical variables. In addition to these numerical data, each patient is also associated with a text document containing intake notes written by nursing staff at the time of ICU admission.\footnote{For the purposes of this study, all text data were pre-processed to remove formatting, punctuation, and spelling errors. After pre-processing, the final corpus of $N$=2,401 documents contained a vocabulary of 14,266 unique terms, with each document containing between two and 861 terms.} The primary outcome in this study was 28-day mortality from the time of ICU admission.

Because the treatment in this study was not randomly assigned to patients, it is possible that patients in the treatment and control groups may differ systematically in ways that affect both their assignment to treatment versus control and their 28-day mortality. For instance, patients who are in critical condition when admitted into the ICU may die before treatment with a TTE has been considered. Similarly, patients whose health conditions quickly improve after admission may be just as quickly discharged. Therefore, in order to obtain unbiased estimates of the effects of TTE on patient mortality, it is important to identify and appropriately adjust for any potentially confounding variables such as degree of health at the time of admission. 

We apply two different matching approaches to this data: one that matches patients only on numerical data and ignores the text data, and one that matches patients using both the numerical and text data. In the first procedure, following \cite{feng2018tte}, we match treated and control units using optimal one-to-one matching \citep{hansen2006optimal} on estimated propensity scores\footnote{Estimated propensity scores are calculated by fitting a logistic regression of the indicator for treatment assignment (receipt of TTE) on the observed numerical covariates.}. We enforce a propensity score caliper equal to 0.1 standard deviations of the estimated distribution, which discards any treated units for whom the nearest control unit is not within a suitable distance. In the second approach, we perform optimal one-to-one text matching within propensity score calipers. Intuitively, this procedure works by first, via the calipers, reducing the space of possible treated-control pairings in a way that ensures adequate balance on numerical covariates. 
By then performing text matching within this space to select a specific match given a set of candidate matches all within the calipers, we obtain matched samples that are similar with respect to all observed covariates, including the original observed covariates and any variables that were not recorded during the study but can be estimated by summary measures of the text. 

Identifying the optimal text-matching method here requires careful consideration of how text similarity should be defined and evaluated in this medical context. Here, the ideal text-matching method is one that matches documents on key medical concepts and prognostic factors that could both impact choice of using TTE as well as the outcome (i.e., potential confounders) that are captured within the text data. Unlike in the previous application, these features cannot be reliably evaluated by non-expert human coders due to the domain expertise and familiarity with medical jargon necessary to make comparisons between medical documents. Thus, to perform a systematic evaluation of text matching methods in this study, we adopt an information retrieval approach for comparing medical texts that has been widely applied in the biomedical literature \citep{aronson2001effective,  zeng2007term}. 

\begin{figure}[ht]
     \begin{center}
     \includegraphics[width=0.9\textwidth]{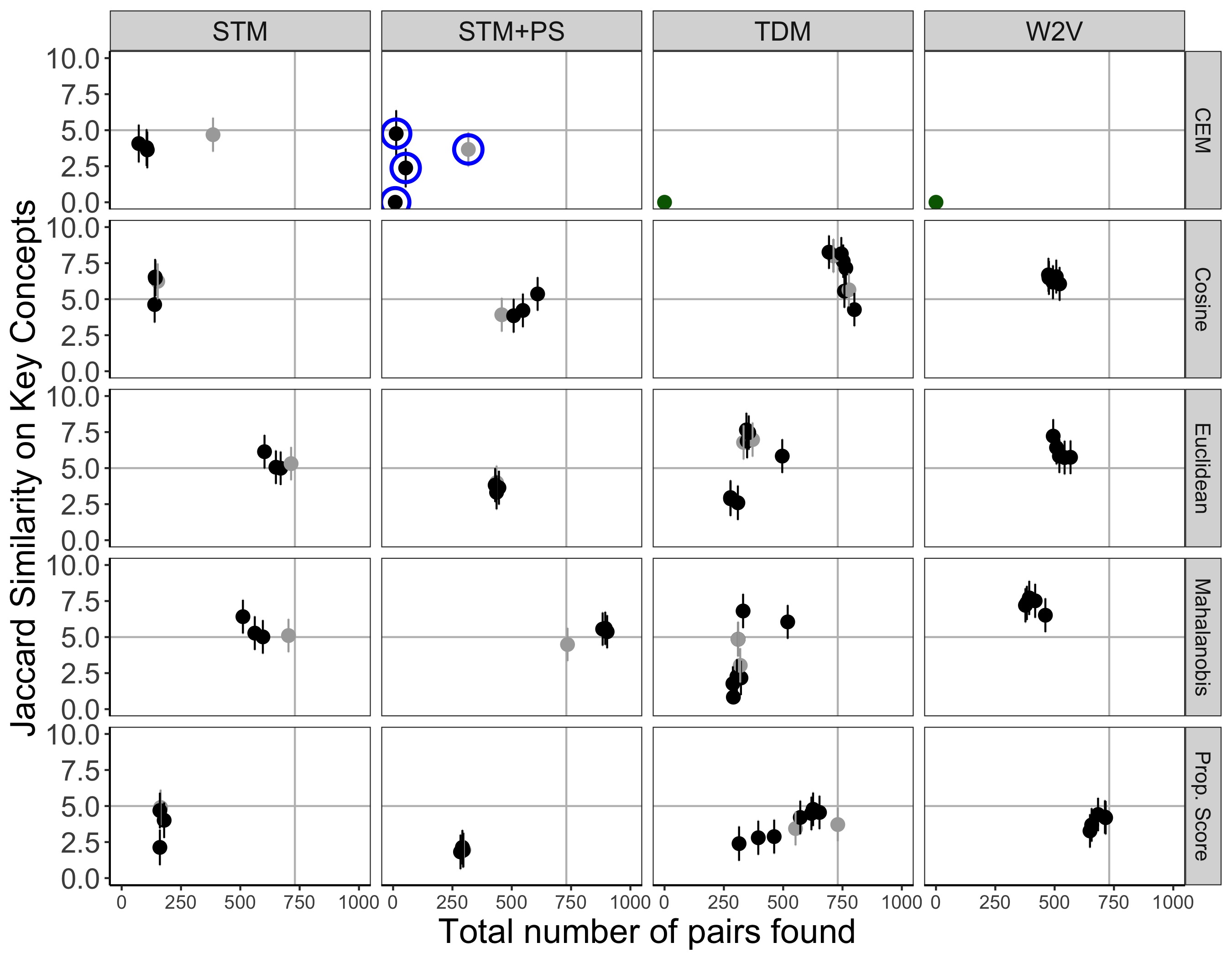}
	\end{center}
	 \caption{Number of matches found versus average pairwise Jaccard similarity for each combination of matching methods. Grey points indicate procedures with extreme reduction in information (e.g., procedures that match on only stop words). Blue circles highlight procedures that use existing state-of-the-art methods for text matching.} 
     \label{fig:mimicJaccard}
\end{figure}

In particular, by consulting with medical professionals, we first obtained a mapping of the texts to a set of clinically meaningful concepts that could be used to characterize ICU patients. Following the approach of \citet{maclean2013identifying}, we then calculated the Jaccard similarity over this mapping between matched pairs of documents as an omnibus measure of match quality. We treat these scores as a working gold standard for this particular application; these scores are based on careful consideration from domain experts who have the medical background required to extract potentially confounding information from this type of nuanced text. When such a mapping is available, the Jaccard similarity metric offers a practical alternative to human evaluation for obtaining estimates of match quality that can be used to compare the relative performance of different matching procedures. However, this metric may not be appropriate for evaluating new texts or for measuring text similarity in other contexts. Figure~\ref{fig:mimicJaccard} shows the average pairwise Jaccard similarity achieved after matching (within propensity score calipers based on the numerical covariates) using each of the 130 text matching specifications described in Section~\ref{sec:methods}. The best-performing procedure matches each treated unit to its nearest control based on the cosine distance calculated over a bounded\footnote{The best identified TDM is bounded to exclude extremely rare and extremely frequent terms, defined operationally as terms that appear in less than four or more than 1000 documents within this corpus.} TDM, where treated units whose nearest control is outside the specified caliper are discarded.

\begin{figure}[ht]
     \begin{center}
     \includegraphics[width=0.9\textwidth]{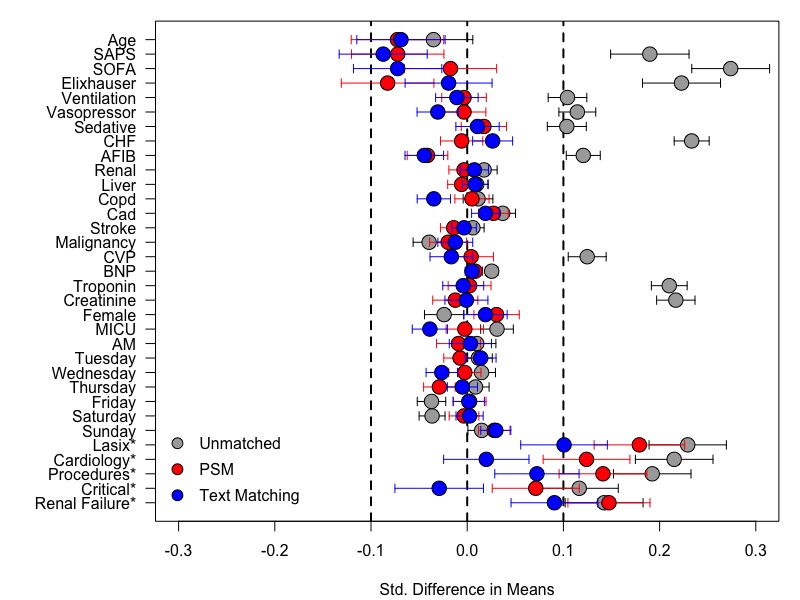}
	\end{center}
	 \caption{Standardized differences in means with 95\% confidence intervals between treatment and control groups on 26 numerical covariates and 5 text-based covariates (denoted by *) before matching (gray), after propensity score matching (red), and after text matching (blue).}
     \label{fig:mimic1}
\end{figure}

Figure~\ref{fig:mimic1} shows the covariate balance between treatment and control groups on both quantitative and text-based covariates before matching, after propensity score matching (PSM) on numeric covariates alone, and after text matching using our preferred method (using cosine distance on a bounded TDM) within propensity score calipers.
Here, each of the five text-based covariates are calculated using summary measures\footnote{The variables \textit{lasix, respiratory, cardiology and critical} are binary variables indicating whether any terms with these words (root terms) were used in the text associated with each patient. The \textit{procedure} variable captures the number of references to medical procedures observed for each patient, and \textit{document length} is defined as the number of words observed for each patient.} based on word-counts from the patient-level text documents.
These variables, according to medical experts consulted on this project, all could indicate potential confounds that could bias estimates of impact if not controlled.
Our general text matching methods do not directly balance these covariates; the improved balance is a consequence of matching on the more general overall distance metric and representation used.

\begin{table}[ht]
\caption{Survival rates for treatment and control groups and estimated treatment effects before and after propensity score matching (PSM) and text matching within propensity score calipers.}
\begin{center}
    \begin{tabular}{|c|c|c|c|c|}
	\hline
    \multirow{2}{*}{Procedure} &
    Effective & 
    \multicolumn{2}{c|}{Survival Rate} & Difference \\ \cline{3-4}
& Sample Size & Treatment & Control & (Std. Error) \\  \hline
Before Matching & 1186 & 72.5\% & 71.2\% & 1.3\% (1.8\%)\\ 
PSM & 807 &72.5\% & 67.7\% & 4.8\% (2.2\%) \\ 
Text Matching  & 894 & 72.5\% & 67.5\% & 5.0\% (2.1\%) \\
\hline
    \end{tabular}
\end{center}
\label{tab:mimic2}
\end{table}

In general, common wisdom (e.g., \citep{imbens2015causal}) is to condition on \textit{all} available data that could indicate potential confounding influences when making inferences using observational data. 
While PSM is able to adequately balance the numerical covariates and some of the text-based covariates most correlated with these numerical measures, it fails to sufficiently adjust for differences between treatment and control groups on a number of potential confounders captured only by the text. 
For instance, both the unmatched data and the matched sample generated using PSM have large imbalances between treatment and control groups on references to Lasix, a medication commonly used to treat congestive heart failure. In the unmatched sample, only 10\% of treated units have documents containing references to this medication compared to 28\% of control units who are associated with the medication. 
Matching on the estimated propensity scores reduces this imbalance only slightly, while cosine matching within propensity score calipers shows a considerable improvement in the balance achieved between treatment groups on this variable. 
Incorporating the text data into the matching procedure leads to similar improvements in balance for the other five text-based variables while also maintaining suitable overall balance on the numerical covariates.

Table~\ref{tab:mimic2} summarizes the survival rates in the treatment and control groups within each matched sample along with the effective sample sizes (i.e., the equivalent number of matched pairs) in the final matched samples. Generally, there appears to be some confounding, with the adjusted impacts being larger than the na\"ive differences.
The matched sample identified using text matching is slightly larger in terms of effective sample size than simple PSM, although they are not significantly different. This increase in effective sample size highlights the efficiency of text matching; when evaluating multiple control units that are eligible matches for a single treated unit in terms of quantitative covariates, the text-based distance offers a more refined measure of pairwise similarity than distances based on the propensity score. Further, when text matching within propensity score calipers, small differences in estimated propensity scores across control units will be offset by any large differences in text. In the present application, this allows for more precise and efficient optimization of the matched sample.

Of course, conducting a matched analysis is rooted in thoughtful design.
In particular, the researcher must decide which variables are important potential confounders, and which are not.
This is especially important when balancing the trade-offs between achieving better balance on some variables at the expense of others.
The purpose of highlighting text matching in this context is to demonstrate how information from the text can also be included in these decisions of what to attend to.
If the text is deemed not informative, then of course it should not be an important consideration with matching.
But, as in this case, if the text is considered to indicate significant aspects of patient condition that should be attended to, the general matching procedures we have discussed can provide a way forward.
And if it is uncertain what is important, then sensitivity checks that focus balance on different groups of variables can further strengthen causal claims in these contexts.

\section{Discussion}
\label{sec:discuss}
In this paper we have made three primary contributions. First, we have provided guidance for constructing different text matching methods and evaluating the match quality of pairs of documents identified using such methods. Second, we empirically evaluated a series of candidate text matching procedures constructed using this framework along with the methods developed in \cite{roberts2018adjusting}. Third, we have applied our methods to a data set of news media in order to engage with a long-standing theoretical debate in political science about the composition of bias in news, and to an observational study evaluating the effectiveness of a medical intervention.

Text matching is widely applicable in the social sciences. \cite{roberts2018adjusting} show how text matching can produce causal estimates in applications such as international religious conflict, government-backed internet censorship, and gender bias in academic publishing. We believe the framework presented in this paper will help expand the scope and usability of text matching even further and will facilitate investigation of text data across a wide variety of disciplines. For instance, the methods described here could enhance state-of-the-art techniques for plagiarism detection and text reuse, techniques that are widely used in political science. By identifying bills that are textually similar to an original legislative proposal, our approach could be used to improve upon work tracking the spread of policy through state legislatures \citep{kroeger2016plagiarizing}; and by comparing social media posts to a matched source article, our methods could detect the dispersion of false news topics through a social network. Secondly, our framework could be used to construct \textit{networks} of lexical similarity, for instance of news sources, politicians, or national constitutions. As well, the metrics we consider for measuring text similarity could themselves resolve measurement problems in cases where lexical \textit{divergence} is the quantity of interest, for example in cases of studying ideological polarization using text data \citep{peterson2018classification}.

We urge, however, that researchers consider how similar their use cases are to ours when extrapolating from results based on our evaluation experiments.
In particular, while cosine distance and TDM-based representations produced high quality results in both of our applied examples, this finding should not be taken as conclusive evidence that these choices are the best in any application of text matching. Further, we emphasize to researchers that the results of our human evaluation experiment depend on the crucial assumption that humans are able to distinguish between textual differences that represent potential confounders, which may bias inferential results if not appropriately controlled for and extraneous differences that are not relevant for the purposes of inference. This assumption may not be plausible in all settings, and we therefore encourage future researchers to conduct their own evaluation studies, especially when using text matching to control for linguistic features other than content similarity, for example stylistic, topic, tone, or semantic similarity. We hope such future evaluations, in connection with this one, will advance our collective understanding of best practices in this important domain.

\newpage
\singlespacing
\bibliographystyle{chicago}
\bibliography{CausalTextBib}
\doublespacing

\newpage
\appendix

\section{Text Representations and Distance Metrics}
\label{app:text}
In Section~\ref{sec:methods} we describe a framework for text matching involving choosing both a text representation and a distance metric; we then briefly outline the options for each. Here we expand that discussion.

\subsection{Choosing a representation}
\label{sec:reps}
To operationalize documents for text matching, we must first represent the corpus in a structured, quantitative form. There are two important properties to consider when constructing a representation for text with the goal of matching. First, the chosen representation should be sufficiently low-dimensional such that it is practical to define and calculate distances between documents. If a representation contains thousands of covariates, calculating even a simple measure of distance may be computationally challenging or may suffer from the curse of dimensionality. Second, the chosen representation should be meaningful; that is, it should capture sufficient information about the corpus so that matches obtained based on this representation will be similar in some clear and interpretable way. As discussed in Section~\ref{sec:background}, text matching is only a useful tool for comparing groups of text documents when the representation defines covariates that contain useful information about systematic differences between the groups.

In this paper, we explore three common types of representations: the term-document matrix (TDM), which favors retaining more information about the text at the cost of dimensionality, statistical topic models, which favor dimension reduction at the potential cost of information, and neural network embeddings, which fall somewhere in between. There are a number of alternative text representations that could also be used to perform matching within our framework, including other representations based on neural networks \citep{bengio2003neural} or those constructed using document embeddings \citep{le2014distributed, dai2015document}, but these are left as a topic for future research.

\subsubsection{Representations based on the term-document matrix}
Perhaps the simplest way to represent a text corpus is as a TDM. Under the common ``bag-of-words'' assumption, the TDM considers two documents identical if they use the same terms with the same frequency, regardless of the ordering of the terms \citep{salton1986introduction}. When matching documents, it is intuitive that documents that use the same set of terms at similar rates should be considered similar, so the TDM provides a natural construction for representing text with the goal of matching. However, the dimensionality of a standard TDM may give rise to computational challenges when calculating pairwise distances between documents in some corpora. There are many dimension-reduction strategies that can be applied to help mitigate this issue including techniques based on matrix rescaling using a scheme such as TF-IDF scoring \citep{salton1991developments}, and techniques for bounding the vocabulary to eliminate extremely rare and/or extremely common terms. However, it should be noted that in large corpora, a bounded and rescaled TDM may still have a dimension in the tens of thousands, setting known to be difficult for matching \citep{roberts2018adjusting}.

\subsubsection{Representations based on statistical topic models}
An alternative representation for text, popular in the text analysis literature, is based on statistical topic models \citep{blei2012probabilistic}, e.g., LDA \citep{blei2003latent} and STM \citep{roberts2016model}. The main argument for matching using a topic-model-based representation of text is that document similarity can adequately be determined by comparing targeted aspects of the text rather than by comparing the use of specific terms. That is, topic-model-based representations imply that two documents are similar if they cover a fixed number of topics at the same rates. Topic models provide an efficient strategy for considerably reducing the dimension of the covariates while retaining all information that is relevant for matching. In contrast to the tens of thousands of covariates typically defined using a representation based on the TDM, representations built using topic models typically contain no more than a few hundred covariates at most. However, consistent estimation of topic proportions is notoriously difficult due to issues with multimodality of these models, which gives rise to a number of issues for applications of matching in practice \citep{roberts2016navigating}.

\subsubsection{Representations based on neural network embeddings}
Mikolov et al. \citep{mikolov2013distributed} introduce a neural network architecture to embed words in an $n-$dimensional space based on its usage and the words which commonly surround it. This architecture has proven remarkably powerful with many intriguing properties. For example, it performs very well in a series of ``linguistic algebra'' tasks, successfully solving questions like ``Japan'' $-$ ``sushi'' $+$ ``Germany'' $=$ ``bratwurst.'' 

\subsubsection{Propensity scores}
When matching in settings with multiple covariates, a common technique is to first perform dimension reduction to project the multivariate covariates into a univariate space. A popular tool used for this purpose is the propensity score, defined as the probability of receiving treatment given the observed covariates \citep{rosenbaum1983central}. Propensity scores summarize all of the covariates into one scalar, and matching is then performed by identifying groups of units with similar values of this score. In practice, propensity scores are generally not known to the researcher and must be estimated using the observed data. When applied to text, propensity scores can be used to further condense the information within a chosen higher-dimensional representation into a summary of only the information that is relevant for determining treatment assignment. Propensity scores representations can be constructed using a quantitative text representation. For example, using STM-based representations or Word2Vec-based representations where dimension of the covariate space is less than the number of documents, standard techniques such as simple logistic regression can be used to estimate propensity scores. To construct propensity score representations over larger a covariate space, such as those typically spanned by a TDM, we use Multinomial Inverse Regression (MNIR; \citealp{taddy2013multinomial}), which provides a novel estimation technique for performing logistic regression of phrase counts from the TDM onto the treatment indicator. After estimating this model, we can calculate a sufficient reduction score that, in principle, will contain all the information from the TDM that is relevant for predicting treatment assignment. Performing a forward regression of the treatment indicator on this sufficient reduction score produces the desired propensity score estimates.

\subsection{Design choices for representations}
Here we discuss a number of design choices that are required for the different representations considered in our study.

\paragraph{TDM-based representations.}
Each of the TDM-based representations is characterized by a bounding scheme, which determines the subset of the vocabulary that will be included in $X$, and a weighting scheme, which determines the numerical rule for how the values of $X$ are measured. We consider standard term-frequency (TF) weighting, TF-IDF weighting, and L2-rescaled TF-IDF weighting. We also consider a number of different screening schemes, including no screening, schemes that eliminate high and low frequency terms, and schemes that consider only high and low frequency terms.

\paragraph{STM-based representations.}
Each STM-based representation is characterized by a fixed number of topics ($K$=10, 30, 50, or 100) and takes one of three distinct forms: 1) the vector of $K$ estimated topic proportions (``S1''), 2) the vector of $K$ estimated topic proportions and the SR score (``S2''), or 3) a coarsened version of the vector of $K$ estimated topic proportions (``S3''). This coarsened representation is constructed using the following procedure. For each document, we first identify the three topics with the largest estimated topic proportions. We retain and standardize these three values and set all remaining $K-3$ topic proportions equal to 0, so that the resulting vector of coarsened topic proportions, $\hat{\theta}^{\star}_i$, contains only three non-zero elements. We then calculate the ``focus'' of each document, denoted by $F_i$, a metric we define as the proportion of topical content that is explained by the three most prominent topics. Focus scores close to one indicate content that is highly concentrated on a small number of topics (e.g., a news article covering health care reform may have nearly 100\% of its content focused on the topics of \textit{health} and \textit{policy}); conversely, focus scores close to zero indicate more general content covering a wide range of topics (e.g., a news article entitled ``The ten events that shaped 2017" may have content spread evenly across ten or more distinct topics). To estimate this score for each document, we take the sum of the raw values of the three non-zero topic proportions identified as above (i.e., $\hat{F}_i = \hat{\theta}_{i[1]} + \hat{\theta}_{i[2]} + \hat{\theta}_{i[3]}$ where $\hat{\theta}_{i[j]}$ is the $j$th order statistic of the vector $\hat{\theta}$). Appending this estimated focus score to the coarsened topic proportion vector produces the final $(K+1)$-dimensional representation.

\paragraph{TIRM representations.}
The TIRM procedure of \cite{roberts2018adjusting} uses an STM-based representation with an additional representation based on document-level propensity scores estimated using the STM framework. These separate representations are then combined within the TIRM procedure using a CEM distance. Each variant of the TIRM procedure considered in this paper is characterized by a fixed number of topics and a set coarsening level (2 bins, 3 bins, or 4 bins). 

\paragraph{Word Embedding representations.}
Google and Stanford University have produced a variety of pre-trained word embedding models. Google's GoogleNews model, where each word vector is length 300 using a corpus of 100 billion words, draws from the entire corpus of Google News; this corpus is therefore extremely well-suited to our analysis. As well, we consider several of Stanford's GloVe embeddings \citep{pennington2014glove}. In particular, we employ their models with word vectors of length 50, 100, 200, and 300. For each of these five embeddings, we produce document-level vectors by taking the weighted average of all word vectors in a document \citep{kusner2015word}.

\subsection{Defining a distance metric} 
After a representation is chosen, applying this representation to the corpus generates a finite set of numerical covariate values associated with each document (i.e., $X_i$ denotes the covariates observed for document $i$ for all $i=1,\ldots,N$). The next step in the matching procedure concerns how to use these covariate values to quantify the similarity between two documents. There are two main classes of distance metrics. Exact and coarsened exact distances regard distances as binary: the distance between two units is either zero or infinity, and two units are eligible to be matched only if the distance between them is equal to zero. Alternatively, continuous distance metrics define distance on a continuum, and matching typically proceeds by identifying pairs of units for whom the calculated distance is within some allowable threshold (``caliper'').  

\subsubsection{Exact and coarsened exact distances}
The exact distance is defined as:
\[D_{ij} = \begin{cases} 0, &\mbox{if } X_i=X_j \\ \infty, &\mbox{otherwise}. \end{cases}\] Matching over this metric (exact matching) generates pairs of documents between treatment and control groups that match exactly on every covariate. Although this is the ideal, exact matching is typically not possible in practice with more than a few covariates. A more flexible metric can be defined by first coarsening the covariate values into ``substantively indistinguishable'' bins, then using exact distance within these bins \citep{iacus2012causal}. For example, using a topic-model-based representation, one might define a coarsening rule such that documents will be matched if they share the same primary topic (i.e., if the topic with the maximum estimated topic proportion among the $K$ topics is the same for both documents). \cite{roberts2018adjusting} advocates using CEM for matching documents based on a representation built using an STM, but, in principle, this technique can also be used with TDM-based representations. For example, one might coarsen the term counts of a TDM into binary values indicating whether each term in the vocabulary is used within each document. Though it is possible in principle, coarsening does not scale well with the dimension of the covariates and so may not be practical for matching with TDM-based representations. This type of distance specification may also create sensitivities in the matching procedure, since even minor changes in the coarsening rules can dramatically impact the resulting matched samples.

\subsubsection{Continuous distances}
\label{sec:subsec:approximate}
Various continuous distance metrics can be used for matching, including linear distances based on the (estimated) propensity score or best linear discriminant \citep{rosenbaum1983central}, multivariate metrics such as the Mahalanobis metric \citep{rubin1973matching}, or combined metrics, such as methods that match on the Mahalanobis metric within propensity score calipers \citep{rosenbaum1985constructing}. When matching on covariates defined by text data, care must be taken to define a metric that appropriately captures the complexities of text. For instance, linear distance metrics such as Euclidean distance may often fail to capture information about the relative importance of different covariates. To make this more clear, consider two pairs of documents containing the texts: {``obama spoke'', ``obama wrote''} and {`he spoke'', ``he wrote''}. Under a TDM-based representation, the Euclidean distances between units in each of these pairs are equal; however, the first pair of documents is intuitively more similar than the second, since the term ``obama'' contains more information about the content of the documents than the term ``he''.  Similarly, the Euclidean distance between the pair documents {``obama spoke'', ``obama obama''} is equivalent to the distance between the pair {``obama spoke'', ``he wrote''}, since by this metric distance increases linearly with differences in term frequencies. These issues also arise when using linear distance metrics with topic-model-based representations.

A metric that is less vulnerable to these complications is Mahalanobis distance, which defines the between documents $i$ and $j$ as $D_{ij} = (X_i-X_i)^T\Sigma^{-1}(X_i-X_j)$, where $\Sigma$ is the variance-covariance matrix of the covariates $X$. This is essentially a normalized Euclidean distance, which weights covariates according to their relative influence on the total variation across all documents in the corpus. Calculating Mahalanobis distance is practical for lower-dimensional representations, but because the matrix inversion does not scale well with the dimension of $X$, it may not be computationally feasible for matching using larger, TDM-based representations.

An alternative metric, which can be efficiently computed using representations defined over thousands of covariates, is cosine distance. Cosine distance measures the cosine of the angle between two documents in a vector space:
\[D_{ij} = 1-\frac{\sum X_iX_j}{\sqrt{\sum X_i^2}\sqrt{\sum X_j^2}}.\]
Cosine distance is commonly used for determining text similarity in fields such as informational retrieval and is an appealing choice for matching because, irrespective of the dimension of the representation, it captures interpretable overall differences in covariate values (e.g., a cosine distance of one corresponds to a 90 degree angle between documents, suggesting no similarity and no shared vocabulary). In general, the utility of a particular continuous distance metric will largely depend on the distribution that is induced on the covariates through the representation.

\subsubsection{Calipers and combinations of metrics}
When pruning treated units is acceptable, exact and coarsened exact matching methods have the desirable property that the balance that will be achieved between matched samples is established a-priori. Treated units for whom there is at least one exact or coarsened exact match in the control group are matched, and all other treated units are dropped. On the other hand, matching with a continuous distance metric requires tuning after distances have been calculated in order to bound the balance between matched samples. After the distances between all possible pairings of treated and control documents have been calculated, one then chooses a caliper, $D_{max}$, such that any pair of units $i$ and $j$ with distance $D_{ij} > D_{max}$ cannot be matched. Here, when pruning treated units is acceptable, any treated units without at least one potential match are dropped. Calipers are typically specified according to a ``rule of thumb'' that asserts that $D_{max}$ be set equal to the value of 0.25 or 0.5 times the standard deviation of the distribution of distance values over all possible pairs of treated and control units, but in some special cases, the caliper can be chosen to reflect a more interpretable restriction. For example, using the cosine distance metric, one might choose a caliper to bound the maximum allowable angle between matched documents.

\subsection{Text as covariates and outcomes}\label{app:covar}
The procedure described in Section~\ref{sec:methods} is relatively straightforward to apply in studies where text enters the problem only through the covariates. However, in more complicated settings where both the covariates and one or more outcomes are defined by features of text, additional steps may be necessary to ensure these components are adequately separated. 

In practice it is generally recommended that outcome data be removed from the dataset before beginning the matching process to preclude even the appearance of ``fishing,'' whereby a researcher selects a matching procedure or a particular matched sample that leads to a desirable result \citep{rubin2007design}. However, this may not be possible when evaluating a text corpus, since both the covariates and outcome may often be latent features of the text \citep{egami2017make}. For instance, suppose we are interested in comparing the level of positive sentiment within articles based on the gender of the authors. One can imagine that news articles that report incidences of crime will typically reflect lower levels of positive sentiment than articles reporting on holiday activities, regardless of the gender of the reporter. Thus, we might like to match articles between male and female reporters based on their topical content and then compare the sentiment expressed within these matched samples. Here, we must extract both the set of covariates that will be used for matching (i.e., topical content) and the outcome (level of positive sentiment) from the same observed text. Because these different components may often be related, measuring both using the same data poses two important challenges for causal inference: first, it requires that the researcher use the observed data to posit a model on the ``post-treatment'' outcome, and, second, measurement of the covariates creates potential for fishing. In particular, suppose that positive sentiment is defined for each document as the number of times terms such as ``happy'' are used within that document (standardized by each document's length). Suppose also that we use the entire vocabulary to measure covariate values for each document (e.g., using a statistical topic model). In this scenario, matching on topical content is likely to produce matches that have similar rates of usage of the term ``happy'' (in addition to having similar rates of usage of other terms), which may actually diminish our ability to detect differences in sentiment. 

To address this issue, we recommend that researchers interested in inference in these settings define the covariates and outcome over a particular representation, or set of distinct representations, such that measurement of the outcome can be performed independently of the measurement of covariates. For example, one might measure the covariates using a representation of text defined over only nouns, and separately, measure outcome values using a representation defined over only adjectives. Or, continuing the previous example, one might divide the vocabulary into distinct subsets of terms, where one subset is used to measure topical content and the other is used to measure positive sentiment. In settings where the chosen representation of the text must be inferred from the observed data (e.g., topic-model-based representations), cross-validation techniques can also be employed, as described in \cite{egami2017make}. For instance, one might randomly divide the corpus into training set and test set, where the training set is used to build a model for the representation, and this model is then applied to the test set to obtain covariate values that will be used in the matching procedure.

\newpage
\section{Index of representations evaluated}
\label{app:index}

\begin{table}[ht]
\label{indexofreps}
\caption{Specification of the 26 representations considered}
\begin{center}
    \begin{tabular}{|c|c|c|c|}
	\hline
    Type & Name  & Description & Dimension \\ \hline
    \multirow{9}{*}{TDM} & T1    & TF Bounded from 4-1000 & 10726  \\
          & T2    & TF-IDF Bounded from 4-1000 & 10726  \\
          & T3    & TF-IDF Bounded from 4-100 & 9413 \\
          & T4    & TF-IDF Bounded from 4-10 & 4879 \\
          & T5    & TF-IDF Bounded from 10-500 & 6000 \\
          & T6    & TF-IDF Bounded from 500-1000 & 154 \\
          & T7    & L2 Rescaled TF-IDF Bounded from 4-1000 & 10726 \\
          & T8    & TF on unbounded TDM & 34397 \\
          & T9    & TF-IDF on unbounded TDM & 34397 \\ \hline
    \multirow{12}{*}{STM} & S1-10 & STM on 10 Topics & 10 \\
          & S2-10 & 10 Topics + estimated sufficient reduction & 11 \\
          & S3-10 & 10 Topics, top 3 topics + focus & 11 \\
          & S1-30 & 30 Topics & 30 \\
          & S2-30 & 30 Topics + estimated sufficient reduction & 31 \\
          & S3-30 & 30 Topics, top 3 topics + focus & 31 \\
          & S1-50 & 50 Topics & 50 \\
          & S2-50 & 50 Topics + estimated sufficient reduction & 51 \\
          & S3-50 & 50 Topics, top 3 topics + focus & 51 \\
          & S1-100 & 100 Topics & 100 \\
          & S2-100 & 100 Topics + estimated sufficient reduction & 101 \\
          & S3-100 & 100 Topics, top 3 topics + focus & 101 \\
\hline
    \multirow{5}{*}{Word2Vec} & W1 & Word embedding of dimension 50 (Google) & 50 \\
          & W2 & Word embedding of dimension 100 (Google) & 100 \\
          & W3 & Word embedding of dimension 200 (Google) & 200 \\
          & W4 & Word embedding of dimension 300 (Google) & 300 \\
          & W5 & Word embedding of dimension 300  & 300 \\
\hline
    \end{tabular}
\end{center}
\end{table}

\newpage

\section{Survey used in human evaluation experiment}
\label{app:surveysnap}
\begin{figure}[ht]
	 \caption{The survey landing page informed participants about the nature of the task.}
     \begin{center}
\includegraphics[width=1.05\textwidth]{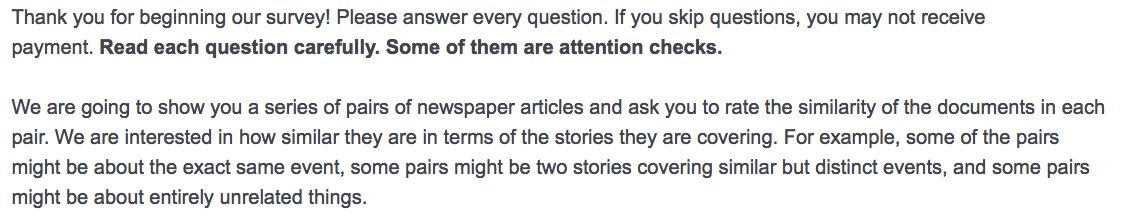}
	\end{center}
     \label{fig:survey1}
\end{figure}

\begin{figure}[ht]
	 \caption{After enrolling in the experiment, participants were presented with a scoring rubric to use as a guide for determining the similarity of a pair of documents.}
     \begin{center}
\includegraphics[width=1.05\textwidth]{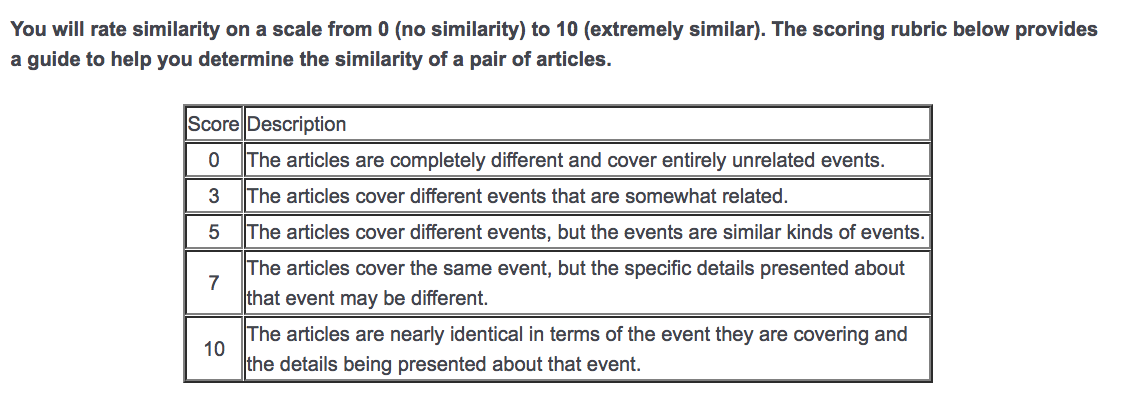}
	\end{center}
     \label{fig:survey2}
\end{figure}

\begin{figure}[ht]
	 \caption{In the first training task of the survey, participants were ask to read and score a pair of articles and were then informed that the anticipated score for this pair was zero. Specifically, they were told ``We think these articles' similarity is 0 out of 10. The first article is related to macaroni and cheese, while the second article is about a murder trial."}
     \begin{center}
\includegraphics[width=1.0\textwidth]{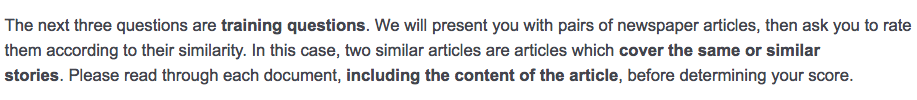}
\includegraphics[width=1.0\textwidth]{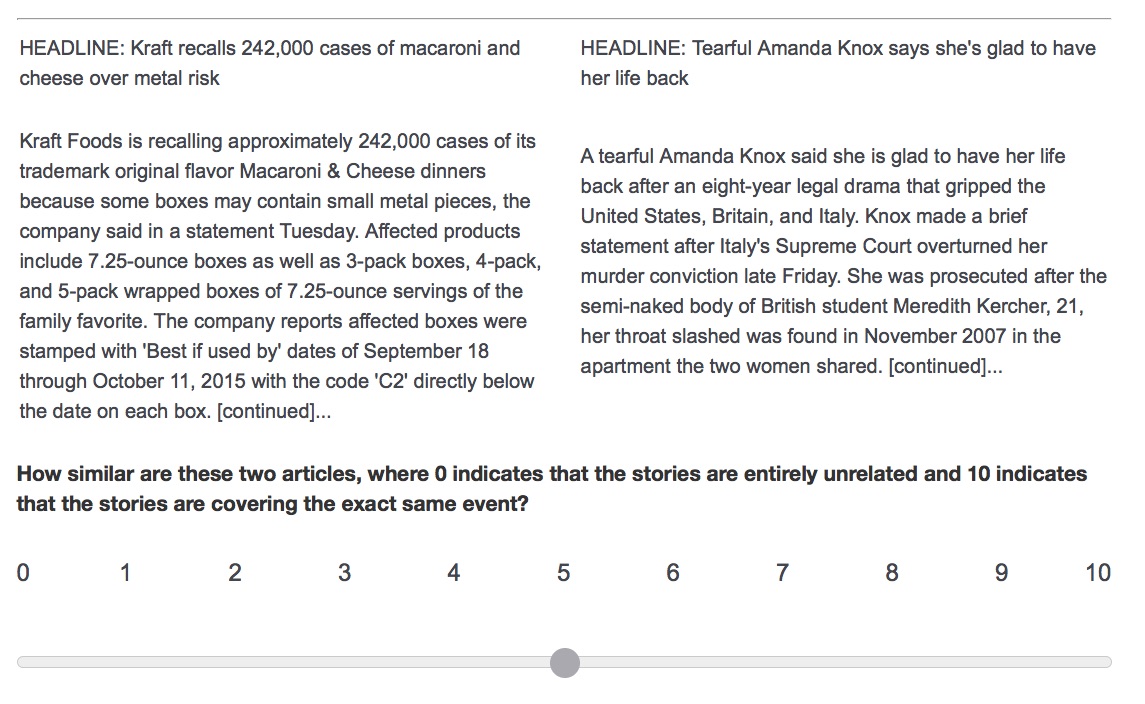}
	\end{center}
     \label{fig:survey3}
\end{figure}

The figures below show snapshots of different components of the survey as they were presented to participants in each of our human evaluation experiments. In particular, Figure~\ref{fig:survey1} shows the survey landing page, where participants were informed about the nature of the task. Participants were then presented with the scoring rubric shown in Figure~\ref{fig:survey2} and completed a series of training tasks as depicted in Figure~\ref{fig:survey3}.

\newpage

\newpage
\section{Sensitivity of match quality scores to the population of respondents}
\label{app:reliability}

\begin{figure}[h]
	\begin{center}
		\includegraphics[width=0.9\textwidth]{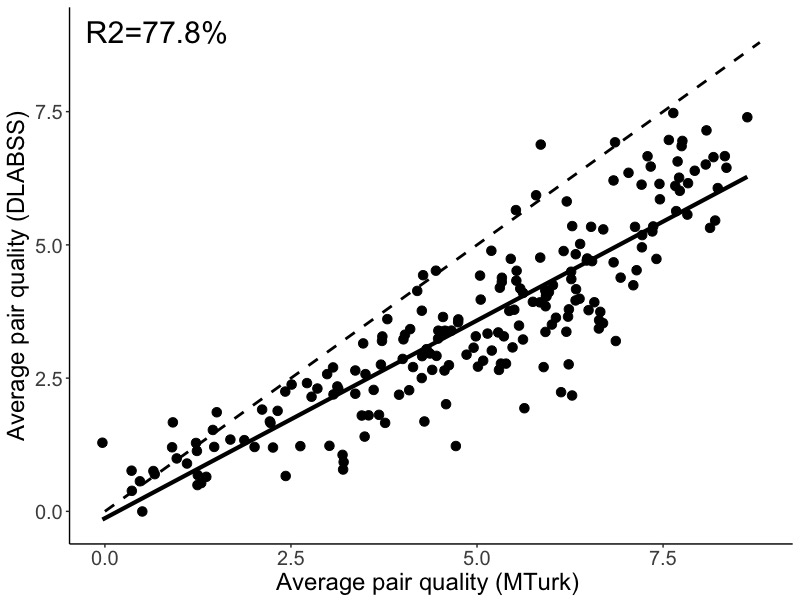}
	\end{center}
	\caption{The strong linear relationship between the average match quality scores for 200 pairs of articles evaluated in two separate pilot studies (solid line) compared to a perfect fit (dotted line) suggests that the survey produces consistent results across samples, when averaged across multiple respondents.}
	\label{fig:pilot}
\end{figure}

To determine the generalizability of the match quality ratings obtained from our survey experiment, we compare two identical pilot surveys using respondents from two distinct populations. The first pilot survey was administered through Mechanical Turk, and the second pilot was administered through the Digital Laboratory for the Social Sciences \citep{dlabss}. For each survey, respondents were asked to read and evaluate ten paired articles, including one attention check and one anchoring question. Each respondent was randomly assigned to evaluate eight matched pairs from a sample of 200, where this pilot sample was generated using the same weighted sampling scheme described above. Figure~\ref{fig:pilot} shows the average match quality scores for each of the 200 matched pairs evaluated based on sample of 337 respondents from Mechanical Turk and 226 respondents from DLABSS. The large correlation between average matched quality scores across samples ($\rho$=0.88) suggests that our survey is a useful instrument for generating consistent average ratings of match quality across diverse populations of respondents. In particular, even though individual conceptions of match quality may differ across respondents, the average of these conceptions both appears to meaningfully separate the pairs of documents and to be stable across at least two different populations.

\newpage

\newpage

\section{Technical details of the evaluation of match quality of pairs of news articles}
\label{app:math}
In this section we more fully describe the design and analysis of the human evaluation experiment for the newspaper matching example.
We start by discussing how we generated our sampling strategy and weights, and then discuss how we used model-assisted survey sampling to estimate average match quality for the different methods along with associated uncertainty.

\subsection{Details of the sampling design}

The study presented in this paper is in fact a replication study as our initial study did not directly assess all procedures considered (in particular, we did not initially evaluate the Word2Vec procedures).
We therefore designed our second study to both directly extend our findings, verify the prior results, and further investigate the predictive accuracy of our models to out-of-sample pairs.
In order to achieve this, we designed a sampling scheme that has three components: (1) we sampled 4 pairs from each procedure considered, (2) we directly sampled pairs that were previously evaluated to assess the stability of the evaluation process, and (3) we sampled pairs not selected by any method to examine differences between  selected and non-selected pairs.
The first stage sampled pairs with weights based on the predicted quality of the pairs in order to sample predicted high-quality pairs more heavily.
We used the prior study's fit predictive model to generate these predictions.
The second and third stage sampled a fixed number of pairs within each tier of quality (from 0 to 8+) to see the full range of pair qualities in our sample (simple random sampling would not work since the vast majority of pairs are scored as quality 1 or lower).
This overall process resulted in a sample of 505 pairs that fully represents all possible pairs (selected and not).
For each pair we have an initial predicted quality score, a sampling probability $\pi_i$, and an associated sampling weight $w_i \propto 1/\pi_i$.

Because many of the procedures generally select the same high-quality pairs, the sequential sampling of 4 pairs for each procedure tends to give many of the same pairs back.
This is by design, and means that our sample primarily consists of pairs shared by multiple procedures which gives greater precision in estimating these procedures' average quality.
We simply take the unique set of pairs sampled as our final evaluation sample.

We calculate the actual sampling weights of each pair for this scheme using simulation.
In particular, we conduct our sampling scheme 100,000 times and calculate how often each pair is selected into the sample.
These provide (up to monte carlo error) the true selection probabilities $\pi_i$; inverting them provides the true sampling weights $w_i$.
For the out-of-matched pairs sampling stage (3), we averaged these final weights across groups of pairs that all have the same probability of selection to increase precision.

The stage (1) sampling scheme intentionally induces selection bias into the sample by discouraging rare pairs, especially singleton pairs, which are expected to be low quality with little variability, in favor of pairs that are identified by multiple matching procedures. 
Regardless, because the sampling probabilities are fixed a-priori, weighted averages of the pairs' match quality gives good estimates of the average quality of the pairs selected by each procedures; this approach is simply classic survey sampling as described in, e.g.,\cite{sarndal2003model}.
All this complexity in the sampling design is to ensure that the sample evaluated is targeted to give information on as many procedures as possible, a difficult task when evaluating 130 procedures with a sample size of about 500.

\subsection{Estimating pair and procedure quality.}
Let $u_{t,c}$ denote a potential pairing of treatment and control documents, where $t$ is the index of the treated unit and $c$ is the index of the control unit. In our evaluation study, $t=1,\ldots,1565$ and $c=1,\ldots,1796$. 
For matching procedure $j$, let $\mathcal{R}_j$ denote the set of $n_j$ matched pairs of articles identified using procedure $j$. 
The set of all unique pairs selected by any of the $J$ procedures considered in the evaluation experiment, denoted $\mathcal{R}$, is defined by the union of these subsets:
\[\mathcal{R} = \cup_{j=1}^{J} \mathcal{R}_j. \]
We index the pairs with $i = 1,\ldots,N$.

The frequency of how often each pair $u_i$ in $\mathcal{R}$ was selected by a procedure is:
\[F_i = \sum_{j=1}^{J} 1\{u_i \in \mathcal{R}_j\}, \]
\noindent where $1\{i \in \mathcal{R}_j\}$ is an indicator variable taking value 1 if pair $u_i$ is identified using matching procedure $j$ and 0 otherwise. 

From the human evaluation, we, for each element $i$ of $\mathcal{S}$, where $\mathcal{S}$ is the set of all sampled pairs, observe $m_i$ similarity ratings, $q^{obs}_{i,1},\ldots,q^{obs}_{i,m_i}$ where $q^{obs}_{i,\cdot} \in [0,10]$. 
We estimate the match quality for each evaluated pair $i$ using the average of observed ratings for that pair, $\bar{q}^{obs}_i$.\footnote{We also explored modeling these ratings to account for rater effects and variable number of ratings per question, but as the results were essentially unchanged, elected to use the simple averages.}

We wish to estimate, for each procedure, the finite-population quantities of the average true quality of the pairs selected.
In particular, if we let $q_i$ be the average quality score score we would see if we had an arbitrarily large number of human respondents evaluate that pair, our targets of inference are, for each procedure $j$,
\[ Q_j = \frac{1}{N_j} \sum_{{u_i \in \mathcal{R}_j }} q_i .\]
The $Q_j$ are population quantities of how the matching procedure did in the specific context considered.
This estimand does not necessarily take into account how the methods would perform on other corpora, even ones similar to this one.

To estimate $Q_j$ for any matching procedure $j$ in our evaluation we use a weighted average of the match quality estimates across the pairs contained in $\mathcal{R}_j \cap \mathcal{S}$, where weights for each pair are equal to the inverse probability of being sampled:
\begin{equation}
\hat{Q}_{samp,j} = \frac{1}{Z_j} \sum_{u_i \in \mathcal{R}_j } \frac{1}{\pi_i} S_i \bar{q}^{obs}_i \mbox{ with } Z_j = \sum_{u_i \in \mathcal{R}_j } \frac{1}{\pi_i} S_i . \label{eq:raw_quality_estimate}
\end{equation} 
with $S_i$ an indicator of whether pair $i$ was sampled for evaluation, with sampling probability $\pi_i$, and $Z_j$ a normalizing constant.
This is a simple Ha\'jek estimator and is known to have good properties.

Unfortunately, despite the sampling scheme, some of our methods only had a small number of pairs sampled for evaluation.
Estimating the average match quality for such procedures could therefore be fairly imprecise.
We address this by using our model for predicting the match quality of a pair of documents based on different machine measures of similarity to construct model-assisted survey sampling estimators that use the predicted qualities to adjust these estimated average quality scores.
We describe this analysis approach next.

\subsection{Improving the estimates of procedure quality.}
To enhance our predictions of match quality for our procedures, we use a model trained on the pairs in $\mathcal{S}_{pre}$, the sample collected in our initial study, to calculate the predicted match quality, $\hat{q}_i$ for all pairs $i=1,\ldots,N$.
These $\hat{q}_i$ are fixed, and do not depend on the analyzed (i.e., second) random sample.
We can use these predictions to adjust our estimates of the average quality of all pairs for each procedure using survey sampling methods.

In particular, our model adjusted quality for procedure $j$ is
\[ \hat{Q}_{adj,j} = \frac{1}{n_j} \sum_{{u_i \in \mathcal{R}_j }} \hat{q}_i + \frac{1}{Z_j} \sum_{{u_i \in \mathcal{R}_j }} S_i \frac{1}{\pi_i} \left( \bar{q}^{obs}_i - \hat{q}_i \right) \]
Here $\hat{q}_i$ is the predicted quality based on the initial sample.
Note the first term in the above is a fixed constant, not dependent on the sample. 
The second term is random, depending on the sample, and, ignoring the small bias induced by $Z_j$ being random, we see the expected value is
\begin{align*}
\EE{ \hat{Q}_{adj,j} } &\approx \frac{1}{n_j} \sum_{{u_i \in \mathcal{R}_j }}  \hat{q}_i + \frac{1}{\EE{Z_j}} \sum_{{u_i \in \mathcal{R}_j }}\EE{  S_i  \frac{1}{\pi_i} \left( \bar{q}^{obs}_i - \hat{q}_i \right) } \\
&= \frac{1}{n_j} \sum_{{u_i \in \mathcal{R}_j }} \hat{q}_i + \frac{1}{\EE{Z_j}} \sum_{{u_i \in \mathcal{R}_j }} \EE{ S_i } \frac{1}{\pi_i} \left( \EE{ \bar{q}^{obs}_i } - \hat{q}_i \right) \\
&=  \frac{1}{n_j} \sum_{{u_i \in \mathcal{R}_j }} q_i  =  Q_j .
\end{align*} 

This is a \emph{model-adjusted estimate}; the first summation gives the predicted average quality of the method.
The second summation adds an adjustment based on the residuals for the actually sampled and evaluated pairs; this adjustment makes the overall estimate effectively unbiased\footnote{The bias is purely from using a Ha\'jek rather than Horvitz-Thompson estimator, and comes from the normalizing $Z_j$ being a random quantity. It is \emph{not} a function of model misfit or misspecification.} regardless of whether the predictive model is useful, predictive, or even correct.
The more the predictive model aligns with the actual measured values, however, the more precise our estimates will be (as the residuals and adjustment part will get smaller and smaller as predictive accuracy grows).

\subsection{Uncertainty estimation}
Classic survey sampling results allowed us to estimate each procedure's average quality with the estimated qualities of our sampled pairs.
We can also increase the precision of these estimates using model adjustment, using the predicted quality scores to adjust the same by population averaged characteristics.
In both cases, the next step is to obtain appropriate uncertainty estimates (standard errors) for these point estimates.
Unfortunately, the task of appropriately calculating uncertainty in this context for both the raw estimates and the model-adjusted estimates is a surprisingly difficult and subtle problem. 
In particular, while there are classic survey sampling formula that can be used to calculate uncertainty, they are asymptotic and are sensitive to extreme weights (which we have). 
This creates some perverse results (i.e. near zero standard errors) for some of the procedures that only had a few pairs sampled. 
To avoid this we, by instituting a homoscedastic assumption for the error terms, did a parametric simulation to calculate uncertainty in order to work around this problem. 
This procedure captures the variability induced by the varying sample weights and the measurement error due to the human evaluation.
We describe this next.

\paragraph{Uncertainty estimates for the raw quality estimates.}
For the unadjusted quality measures, we estimate uncertainty using the principles of a case-wise bootstrap with some modifications.
In particular, especially for those methods with very few (e.g. 4) sampled pairs, estimating the variability of quality of the pairs via case-wise bootstrap is unreliable unless we pool or partially pool estimates of variability across the different methods. 

To see this consider a hypothetical method with 4 of its pairs sampled, 1 with very high weight due to being a rare pair and 3 with a low weight due to being selected by most methods.
Any bootstrap sample that includes the high weight unit will essentially give an average quality score close to that of the high weight unit.
Even bootstrap samples with multiple draws of the high weight unit will still get nearly that same average quality score since the values of these large elements will all be the same.
Across bootstrap samples, this will give low variability, i.e., seemingly high precision.
It does not take into account the variability of scores we might have actually seen across other units of similar weight.
We address this with the a parametric approach that we describe next.

We first assess the typical variability of the quality scores of pairs within the procedures.
For the unadjusted quality scores of the individual pairs we first calculated an estimate of the standard deviation of scores within a given match method (we did this by calculating the weighted standard deviation of scores).
We then took the median of these values as our measure of within-method variation of pair quality.
We use the median to avoid the impact of the extreme standard deviations due to the methods with small samples of pairs.\footnote{We actually calculated this (pooled) standard deviation a variety of ways and took the largest to be maximally conservative.}

To calculate standard errors for our methods, we then simulated the pair sampling step followed by the scoring of sampled pairs step by first selecting pairs using the original sampling strategy, and then generating pseudo-quality scores with the same variance as we generally saw for pairs selected by a method.
We then calculated the overall pseudo-quality for each of our methods based on these scores and associated sample weights.
Our standard errors are then the standard deviation of these generated overall pseduo-quality scores.

To compare, we also conducted a simple case-wise bootstrap. Here we sampled the evaluated pairs with replacement and calculated each methods' quality score using the bootstrap sample, finally obtaining standard errors using the standard deviation of the resulting values. 
This approach works well for those methods with 10 or more sampled pairs.
Overall, our parametric approach generally produced larger standard errors, which is a mixture of the overall conservatism of our approach and of the aforementioned issue of the na\'ive approach giving small standard errors those methods with few pairs and a few high-weight pairs that dominate the overall quality measure.
We thus report our parametric simulation-based standard errors.

\paragraph{Uncertainty for the model-adjusted approach.}
For the model-adjusted case, we again worried about those methods with few samples having less variability due to small numbers of high weight units giving nearly the same model adjustment with each step. 
We therefore follow the above process, but instead of generating synthetic outcomes we generated synthetic residuals by generating normally distributed noise with variance equal to the variance of the original residuals from our predictive model.
These simulated residual-based standard errors were again conservative when compared to the na\"ive case-wise approach for those procedures with enough selected pairs to make this comparison.

\paragraph{Remarks.} All our uncertainty estimation methods capture the uncertainty in the pair quality evaluation process as the variability of the pairs' quality scores captures both the measurement error and the structural variation of the pairs themselves. 
In our plots, we report the simulation-based standard errors for the model adjusted estimates. 
As noted in the text, the model-adjusted quality scores themselves were generally similar to unadjusted (for the directly evaluated methods where we had both scores), and the differences between the two had no impact on our overall findings.

For methods that we did not initially identify for our human evaluation, we could calculate a predicted quality based on our model of
\[ \hat{Q}_{pred,j} = \frac{1}{n_j} \sum_{{u_i \in \mathcal{R}_j }} \hat{q}_i  .\]
This is extrapolation, however. 
If the new procedure was selecting pairs that systematically were better than predicted, for example, this extrapolation would be biased.
Even if such a new method happened to use some pairs randomly selected for evaluation, we cannot use the survey adjusted $\hat{Q}_{adj,j}$ or raw estimate $\hat{Q}_{samp,j}$ since the pairs \emph{not} in the sampling frame had no chance of selection.
One could create a hybrid estimator by splitting the sample into potentially sampled, but we do not explore that further here.

\subsection{Prior evaluation study details}

As mentioned above, we performed an initial full study on an initial subset of the matching procedures considered (in particular, we did not initially evaluate the Word2Vec procedures).
Overall, this study produced the same results as our final study.

We sampled pairs differently for our initial study.
In particular, we did not have baseline predicted quality scores to calculate sampling weights from.
We therefore, to produce a representative sample of matched articles for evaluation, did not take a sample from each procedure's pair list but instead took a weighted random sample of 500 pairs from the union of these lists, $\mathcal{R}$, with sampling weights roughly proportional to $F_i$, where $F_i$ is the number of times pair $i$ was selected by a procedure.
Because singleton pairs comprised over 75\% of the pairs in $\mathcal{R}$, we further downweighted pairs with $F_i = 1$ by a factor of 5. 
Our overall sampling probabilities for pair $i$ were then
\[ w_i \propto 
\left\{ \begin{array}{l}
0.20  \text{ if $F_i = 1$},\\ 
F_i \text{ otherwise}. 
\end{array} \right.
\]
We then calculated true sampling probabilities and weights via simulation as described above (due to high weights for some pairs and the sampling without replacement these initially weights are not truly proportional to inverse probability of selection).
\newpage

\section{Notes on the sample and unadjusted human experiment results}
\label{app:raw_results}

The final evaluation sample consisted of 33 pairs that were originally evaluated in the initial evaluation, 50 pairs that were not identified by any matching method considered, and 422 pairs that were used by at least one matching method evaluated.
The sampling weights for those pairs that were selected by at least one method ranged from 0.02 to 10.7, with a median of 0.23.
This corresponded to selection probabilities ranging from 1 in 1000 to 77\%.
25\% of the pairs had less than a 1\% chance of being selected.
The very rare pairs tend to come from the propensity score methods that had a large number of low-quality matches.
Across procedures, some had only 4 pairs sampled, and some had up to 100.  The average was 28 pairs.

The standard deviation of quality scores did depend on the sampling weight, with a standard deviation of around 2.5 for low $pi_i$ and 1 for the highest $pi_i$.
On the other hand, the standard deviation of scores for very low and very high predicted qualities was less than 0.5, rising to around 1.6 for pairs predicted to have a quality of 5.
Within a given procedure, scores tended to have a standard deviation of around 2.37, for those procedures with 10 or more pairs sampled.
If we look across all procedures the median decreases markedly due to poor estimates for small sample sizes.
We used 2.37 in our simulation.

For the residual scores, residuals had a lower standard deviation near the endpoints (due to truncation) and peaked at around 1.6 for the middle scores.
We therefore use a residual standard deviation of 1.6 in our simulations to calculate our standard errors, which will be generally conservative.
Even with this conservative approximation, we are explaining 55\% of our variation with our predictive score.

\begin{figure}[h]
	\begin{center}
		\includegraphics[width=0.9\textwidth]{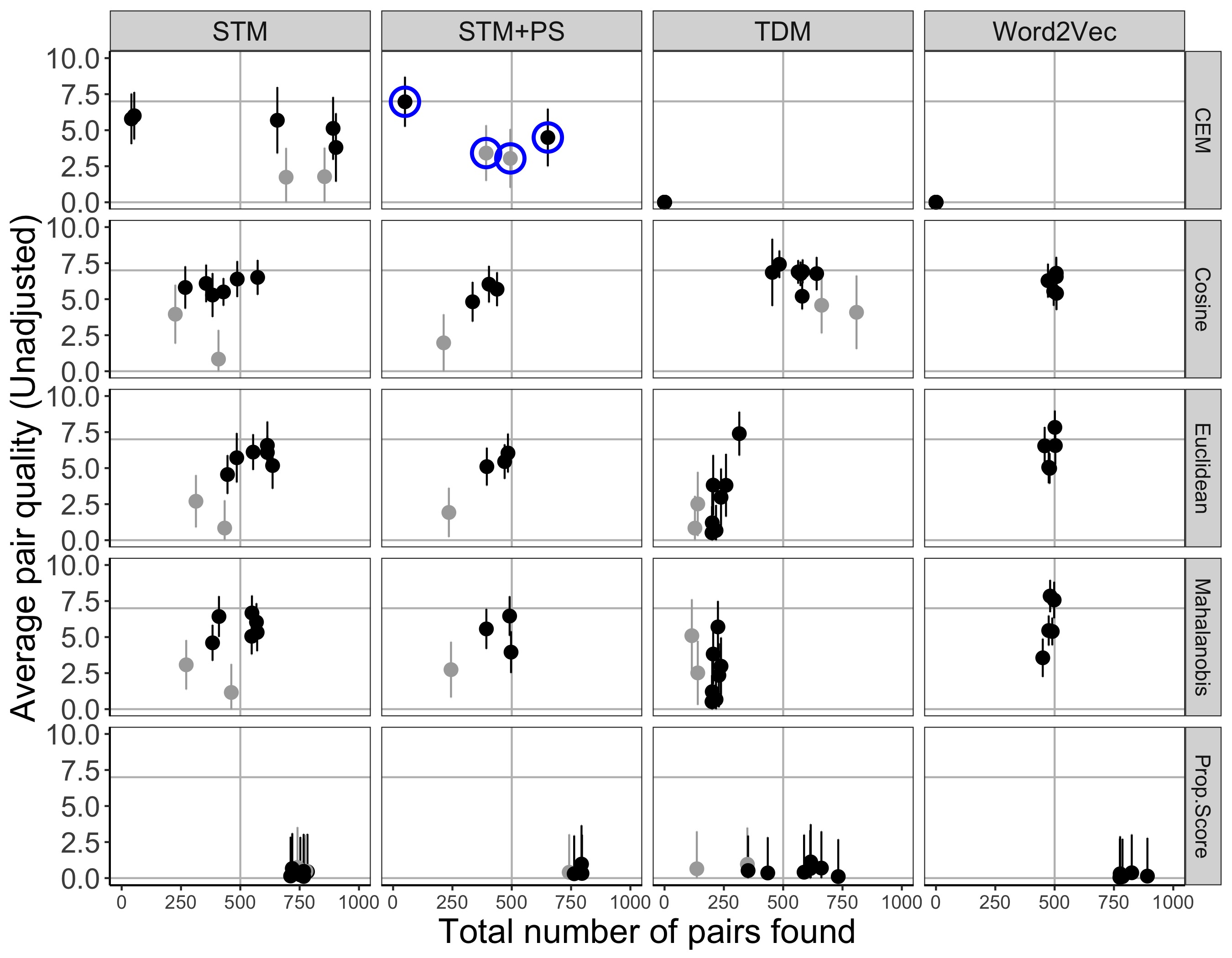}
	\end{center}
	\caption{Number of matches found versus estimated (unadjusted) average match quality scores for each combination of matching methods. Grey points indicate procedures with extreme reduction in information (e.g., procedures that match on only stop words). Blue circles highlight procedures that use existing state-of-the-art methods for text matching. }
	\label{fig:qualitySamp}
\end{figure}

Figure~\ref{fig:qualitySamp} shows the simple weighted average match quality of the directly evaluated pairs sampled for each of the 130 procedures considered in the evaluation experiment.
The nominal 95\% confidence intervals are from standard errors calculated from the parametric bootstrap described above.

The standard errors seem small, but some mild calculations suggest they are reasonable.
In particular, with 28 pairs, if the pairs have a standard deviation of about 2, we would expect, roughly a standard error of $2 / \sqrt{28} = 0.38$, which is what we tended to see.
We also point out that we are considering the population of pairs selected by a method as fixed: this is a finite sample inference problem.

\newpage
\section{Template Matching and Sensitivity Analyses for Media Bias Application}
\label{app:sensitivity_media}

\begin{figure}[ht]
     \begin{center}
     \includegraphics[width=0.5\textwidth]{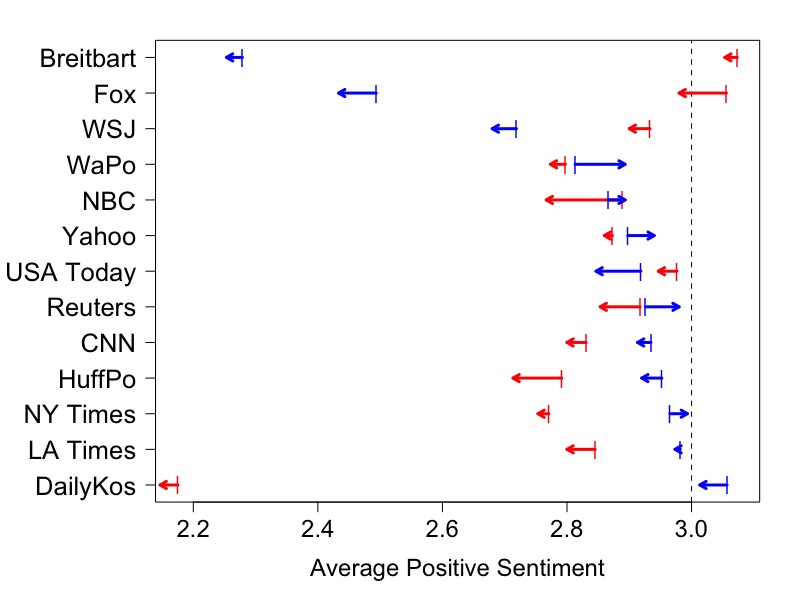}
          \includegraphics[width=0.5\textwidth]{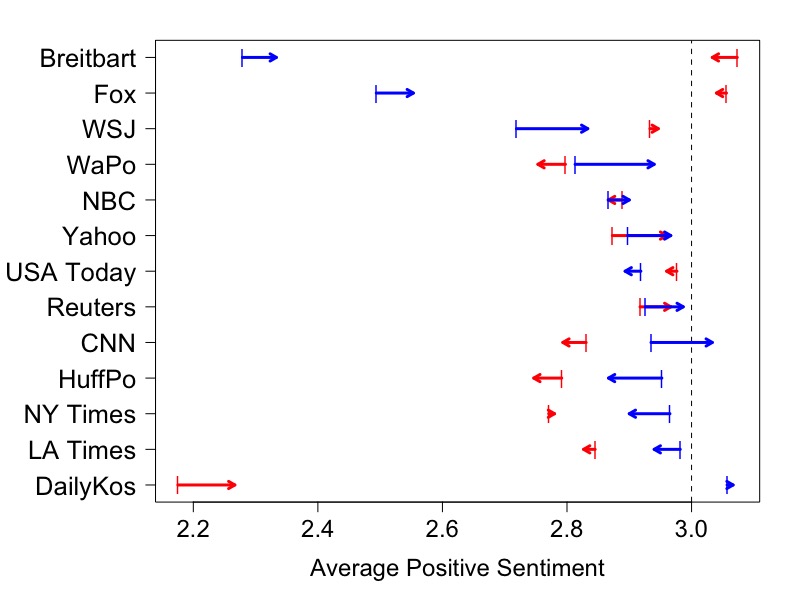}
     \includegraphics[width=0.5\textwidth]{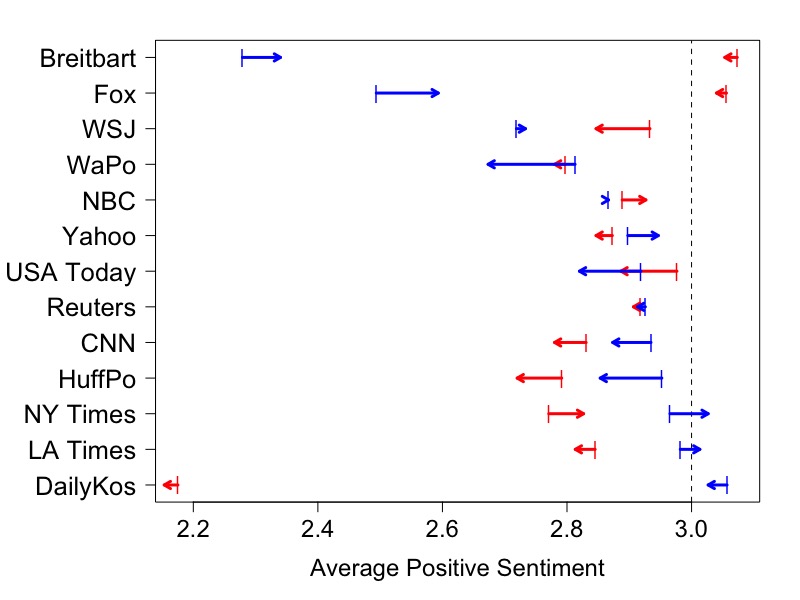}
	\end{center}
	 \caption{Estimates of average favorability toward Democrats (blue) and Republicans (red) for each source both before and after matching using Mahalanobis matching on an STM with 100 topics (top), propensity score matching on an STM with 100 topics (center) and propensity score matching on a TDM (bottom).}
     \label{fig:sensitivity}
\end{figure}

To evaluate the robustness of our findings, we performed a series of sensitivity checks to assess how our results and subsequent conclusions change when using different specifications of the matching procedure. Figure~\ref{fig:sensitivity} shows the results produced by three alternative text matching methods. These robustness checks highlight the importance of the specification of the matching procedure: weaker methods (i.e., methods that produce low quality matches) typically lead to weaker inferences. For example, the results produced from template matching using the Mahalanobis distance metric on a vector of 100 topic proportions show generally smaller changes in average favorability within each source before and after matching than the results shown in Figure~\ref{fig:allmeans}. The null results in this case provide further evidence in support of the claim that text matching is an effective strategy for reducing differences in the observed biases across news sources that are due to topic selection.

\begin{figure}[ht]
	\begin{center}
		\includegraphics[width=1.0\textwidth]{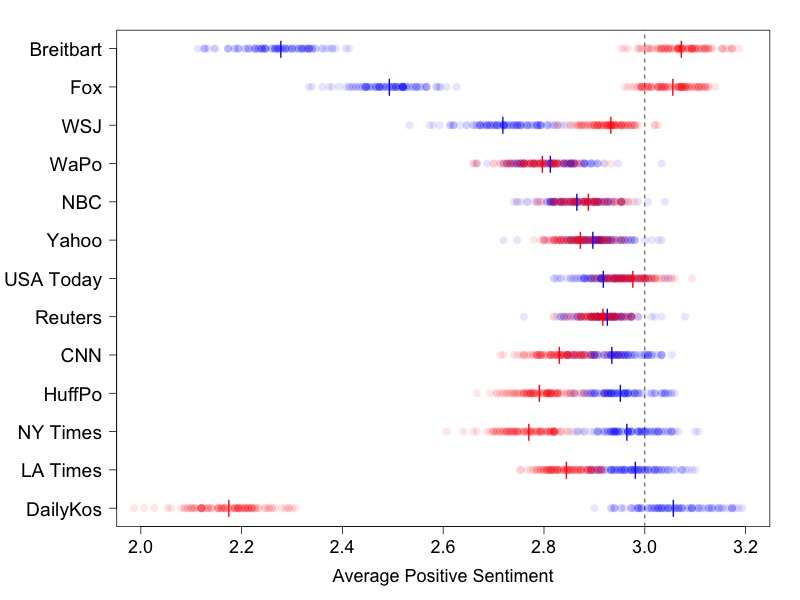}
	\end{center}
	\caption{Estimates of average favorability toward Democrats (blue) and Republicans (red) for each source for 100 iterations of random matching. Blue and red lines represent the average favorability scores before matching for Democrats and Republicans, respectively.}
	\label{fig:sensitivity2}
\end{figure}

As a final robustness check of the results based on our template-matched sample, we performed the following consistency test. First, we randomly generated 10,000 pairs of documents containing 150 randomly selected articles from each news source. In each iteration of random sampling and for each news source, we then calculated the average favorability scores towards Democrats and Republicans within the matched sample. Figure~\ref{fig:sensitivity2} shows the distributions of these favorability scores for each news source after 100 iterations of random matching. Finally, we calculated the total \textit{change} in favorability observed after matching in each iteration, averaged across all 13 sources. More formally, for each iteration $i=1,\ldots,10000$ we calculated the test statistic: \[T_i=\frac{1}{13}\sum_{j=1}^{13} \left(|\hat{Y}_j^{dem}-\hat{Y}_{j,M_i}^{dem}| + |\hat{Y}_j^{rep}-\hat{Y}_{j,M_i}^{rep}|\right),\]
where $\hat{Y}_j^{dem}$ and $\hat{Y}_j^{rep}$ denote the average favorability scores toward democrats and republicans, respectively, for all articles corresponding to source $j$ in the original, unmatched sample. Quantities $\hat{Y}_{j,M_i}^{dem}$ and $\hat{Y}_{j,M_i}^{rep}$ denote the partisan favorability scores averaged across the set of 150 articles from source $j$ that were selected by random matching in iteration $i$. The sampling distribution of this test statistic provides a reference for values of the test statistic that may occur when comparing randomly selected sets of 150 articles across these 13 sources. Therefore, by comparing the value of our observed test statistic based on the results of our template-matching procedure described in Section~\ref{sec:applied} to the randomization distribution defined by $T$, we can estimate the probability that our template-matched results are due to random chance. Our results from this randomization test indicated that template matching on text removes a significant amount of the bias observed across sources that remains after adjusting for differences in topic selection ($p$=0.004). 

\end{document}